\documentclass[twocolumn]{aastex631}
\usepackage{float}

\received{\today}
\revised{\today}
\accepted{\today}

\begin{document}
	
	\title{Ejecta masses in Type Ia Supernovae -- Implications for the Progenitor and the Explosion Scenario\footnote{Based in part on observations obtained with the Hobby-Eberly Telescope (HET), which is a joint project of the University of Texas at Austin, the Pennsylvania State University, Ludwig-Maximillians-Universitaet Muenchen, and Georg-August Universitaet Goettingen. The HET is named in honor of its principal benefactors, William P. Hobby and Robert E. Eberly.}
	}

	\correspondingauthor{Zs\'ofia Bora}
	\email{bora.zsofia@csfk.hun-ren.hu}
	
	\author[0000-0001-6232-9352]{Zs\'ofia Bora}
	\affiliation{HUN-REN CSFK Konkoly Observatory,
		Konkoly Thege M. \'ut 15-17, Budapest, 1121, Hungary}
	\affiliation{ELTE E\"otv\"os Lor\'and University, Institute of Physics and Astronomy, 
		P\'azm\'any P\'eter s\'et\'any 1A, Budapest 1117, Hungary}

	\author[0000-0002-8770-6764]{R\'eka K\"onyves-T\'oth}
	\affiliation{HUN-REN CSFK Konkoly Observatory, Konkoly Thege M. \'ut 15-17, Budapest, 1121, Hungary}
	\affiliation{Department of Experimental Physics, Institute of Physics, University of Szeged, D\'om t\'er 9, Szeged, 6720 Hungary}
	
	\author[0000-0001-8764-7832]{J\'ozsef Vink\'o}
	\affiliation{HUN-REN CSFK Konkoly Observatory, 
		Konkoly Thege M. \'ut 15-17, Budapest, 1121, Hungary}
	\affiliation{Department of Experimental Physics, Institute of Physics, University of Szeged, D\'om t\'er 9, Szeged, 6720 Hungary}
	\affiliation{ELTE E\"otv\"os Lor\'and University, Institute of Physics and Astronomy,  
		P\'azm\'any P\'eter s\'et\'any 1A, Budapest 1117, Hungary}
	\affiliation{Department of Astronomy, University of Texas at Austin,
		2515 Speedway, Stop C1400, Austin, TX, 78712-1205, USA}

	\author{Dominik B\'anhidi}
	\affiliation{Baja Astronomical Observatory of University of Szeged, Szegedi \'ut, Kt. 766, 6500 Baja, Hungary}
	
	\author[0000-0001-9061-2147]{Imre Barna B\'ir\'o}
	\affiliation{Baja Astronomical Observatory of University of Szeged, Szegedi \'ut, Kt. 766, 6500 Baja, Hungary}
	\affiliation{ELKH-SZTE Stellar Astrophysics Research Group, Szegedi \'ut, Kt. 766, 6500 Baja, Hungary}
	
	\author[0000-0002-4924-444X]{K.\ Azalee Bostroem}
	\altaffiliation[]{LSST-DA Catalyst Fellow}
	\affiliation{Steward Observatory, University of Arizona, 933 North Cherry Avenue, Tucson, AZ 85721-0065, USA}
	
	\author[0000-0002-8585-4544]{Attila B\'odi}
	\affiliation{HUN-REN CSFK Konkoly Observatory,
		Konkoly Thege M. \'ut 15-17, Budapest, 1121, Hungary}
	
	\author{Jamison Burke}
	\affiliation{Las Cumbres Observatory, 6740 Cortona Drive, Suite 102, Goleta, CA 93117-5575, USA}
	
	\author{Istv\'an  Cs\'anyi}
	\affiliation{Baja Astronomical Observatory of University of Szeged, Szegedi \'ut, Kt. 766, 6500 Baja, Hungary}
	
	\author[0000-0002-6497-8863]{Borb\'ala Cseh}
	\affiliation{HUN-REN CSFK Konkoly Observatory,
		Konkoly Thege M. \'ut 15-17, Budapest, 1121, Hungary}
	\affiliation{MTA-ELTE Lend{\"u}let "Momentum" Milky Way Research Group, Hungary}
	
	\author[0000-0003-4914-5625]{Joseph Farah}
	\affiliation{Las Cumbres Observatory, 6740 Cortona Drive, Suite 102, Goleta, CA 93117-5575, USA}
	\affiliation{Department of Physics, University of California, Santa Barbara, CA 93106-9530, USA}
	
	\author[0000-0003-3460-0103]{Alexei V. Filippenko}
	\affiliation{Department of Astronomy, University of California, Berkeley, CA 94720-3411, USA}
	
	\author{Tibor Heged\"us}
	\affiliation{Baja Astronomical Observatory of University of Szeged, Szegedi \'ut, Kt. 766, 6500 Baja, Hungary}
	
	\author[0000-0002-1125-9187]{Daichi Hiramatsu}
	\affiliation{Center for Astrophysics \textbar{} Harvard \& Smithsonian, 60 Garden Street, Cambridge, MA 02138-1516, USA}
	\affiliation{The NSF AI Institute for Artificial Intelligence and Fundamental Interactions, USA}
	
	\author{\'Agoston Horti-D\'avid}
	\affiliation{ELTE E\"otv\"os Lor\'and University, Institute of Physics and Astronomy, 
		P\'azm\'any P\'eter s\'et\'any 1A, Budapest 1117, Hungary}
	\affiliation{HUN-REN CSFK Konkoly Observatory,
		Konkoly Thege M. \'ut 15-17, Budapest, 1121, Hungary}

	\author[0000-0003-4253-656X]{D.\ Andrew Howell}
	\affiliation{Las Cumbres Observatory, 6740 Cortona Drive, Suite 102, Goleta, CA 93117-5575, USA}
	\affiliation{Department of Physics, University of California, Santa Barbara, CA 93106-9530, USA}
	
	\author[0000-0001-8738-6011]{Saurabh W. Jha}
	\affiliation{Department of Physics and Astronomy, Rutgers, the State University of New Jersey, 136 Frelinghuysen Road, Piscataway, NJ 08854, USA}

	\author[0000-0002-1663-0707]{Csilla Kalup}
	\affiliation{ELTE E\"otv\"os Lor\'and University, Institute of Physics and Astronomy, 
		P\'azm\'any P\'eter s\'et\'any 1A, Budapest 1117, Hungary}
	\affiliation{HUN-REN CSFK Konkoly Observatory,
		Konkoly Thege M. \'ut 15-17, Budapest, 1121, Hungary}
	
	\author[0000-0002-8813-4884]{M\'at\'e Krezinger}
	\affiliation{ELTE E\"otv\"os Lor\'and University, Institute of Physics and Astronomy, 
		P\'azm\'any P\'eter s\'et\'any 1A, Budapest 1117, Hungary}
	\affiliation{HUN-REN CSFK Konkoly Observatory,
		Konkoly Thege M. \'ut 15-17, Budapest, 1121, Hungary}
	
	\author{Levente Kriskovics}
	\affiliation{HUN-REN CSFK Konkoly Observatory,
		Konkoly Thege M. \'ut 15-17, Budapest, 1121, Hungary}
	
	\author[0000-0001-5807-7893]{Curtis McCully}
	\affiliation{Las Cumbres Observatory, 6740 Cortona Drive, Suite 102, Goleta, CA 93117-5575, USA}
	
	\author[0000-0001-9570-0584]{Megan Newsome}
	\affiliation{Las Cumbres Observatory, 6740 Cortona Drive, Suite 102, Goleta, CA 93117-5575, USA}
	\affiliation{Department of Physics, University of California, Santa Barbara, CA 93106-9530, USA}
	
	\author{Andr\'as Ordasi}
	\affiliation{HUN-REN CSFK Konkoly Observatory,
		Konkoly Thege M. \'ut 15-17, Budapest, 1121, Hungary}
	
	\author[0000-0003-0209-9246]{Estefania Padilla Gonzalez}
	\affiliation{Las Cumbres Observatory, 6740 Cortona Drive, Suite 102, Goleta, CA 93117-5575, USA}
	\affiliation{Department of Physics, University of California, Santa Barbara, CA 93106-9530, USA}
	
	\author[0000-0001-5449-2467]{Andr\'as P\'al}
	\affiliation{HUN-REN CSFK Konkoly Observatory,
		Konkoly Thege M. \'ut 15-17, Budapest, 1121, Hungary}
	
	\author[0000-0002-7472-1279]{Craig Pellegrino}
	\affiliation{Las Cumbres Observatory, 6740 Cortona Drive, Suite 102, Goleta, CA 93117-5575, USA}
	\affiliation{Department of Physics, University of California, Santa Barbara, CA 93106-9530, USA}
	
	\author[0000-0002-3658-2175]{B\'alint Seli}
	\affiliation{ELTE E\"otv\"os Lor\'and University, Institute of Physics and Astronomy, 
		P\'azm\'any P\'eter s\'et\'any 1A, Budapest 1117, Hungary}
	\affiliation{HUN-REN CSFK Konkoly Observatory,
		Konkoly Thege M. \'ut 15-17, Budapest, 1121, Hungary}
	
	\author[0000-0001-7806-2883]{\'Ad\'am S\'odor}
	\affiliation{HUN-REN CSFK Konkoly Observatory,
		Konkoly Thege M. \'ut 15-17, Budapest, 1121, Hungary}
	
	\author[0000-0001-9830-3509]{Zs\'ofia Marianna Szab\'o}
	\affiliation{HUN-REN CSFK Konkoly Observatory,
		Konkoly Thege M. \'ut 15-17, Budapest, 1121, Hungary}
	\affiliation{Scottish Universities Physics Alliance (SUPA), School of Physics and Astronomy, University of St Andrews, North Haugh, St Andrews, KY16 9SS, UK}
	\affiliation{Max-Planck-Institut für Radioastronomie, Auf dem Hügel 69, 53121 Bonn, Germany}

	\author{Norton O. Szab\'o}
	\affiliation{ELTE E\"otv\"os Lor\'and University, Institute of Physics and Astronomy, 
		P\'azm\'any P\'eter s\'et\'any 1A, Budapest 1117, Hungary}
	\affiliation{HUN-REN CSFK Konkoly Observatory,
		Konkoly Thege M. \'ut 15-17, Budapest, 1121, Hungary}
	
	\author[0000-0002-1698-605X]{R\'obert Szak\'ats}
	\affiliation{HUN-REN CSFK Konkoly Observatory,
		Konkoly Thege M. \'ut 15-17, Budapest, 1121, Hungary}
	
	\author[0000-0003-4610-1117]{Tam\'as Szalai}
	\affiliation{Department of Experimental Physics, Institute of Physics, University of Szeged, D\'om t\'er 9, Szeged, 6720 Hungary}
	
	\author{P\'eter Sz\'ekely}
	\affiliation{Department of Experimental Physics, Institute of Physics, University of Szeged, D\'om t\'er 9, Szeged, 6720 Hungary}
	
	\author[0000-0003-0794-5982]{Giacomo Terreran}
	\affiliation{Las Cumbres Observatory, 6740 Cortona Drive, Suite 102, Goleta, CA 93117-5575, USA}
	
	\author{V\'azsony Varga}
	\affiliation{ELTE E\"otv\"os Lor\'and University, Institute of Physics and Astronomy, 
		P\'azm\'any P\'eter s\'et\'any 1A, Budapest 1117, Hungary}
	\affiliation{HUN-REN CSFK Konkoly Observatory, 
		Konkoly Thege M. \'ut 15-17, Budapest, 1121, Hungary}
	
	\author[0000-0002-6471-8607]{Kriszti\'an Vida}
	\affiliation{HUN-REN CSFK Konkoly Observatory,
		Konkoly Thege M. \'ut 15-17, Budapest, 1121, Hungary}
	
	\author[0000-0002-7334-2357]{Xiaofeng Wang}
	\affiliation{Physics Department and Tsinghua Center for Astrophysics (THCA), Tsinghua University, Beijing, China}
	\affiliation{Beijing Planetarium, Beijing Academy of Science and Technology, Beijing, China}
	
	\author[0000-0003-1349-6538]{J. Craig Wheeler}
	\affiliation{Department of Astronomy, University of Texas at Austin,
		2515 Speedway, Stop C1400, Austin, TX, 78712-1205, USA}

	\begin{abstract}
		The progenitor system(s) as well as the explosion mechanism(s) of thermonuclear (Type Ia) supernovae are long-standing issues in astrophysics. Here we present ejecta masses and other physical parameters for 28 recent Type Ia supernovae inferred from multiband photometric and optical spectroscopic data. Our results confirm that the majority of SNe Ia show {\it observable} ejecta masses below the Chandrasekhar-limit (having a mean $M_{\rm ej} \approx 1.1 \pm 0.3$ M$_\odot$), consistent with the predictions of recent sub-M$_{\rm Ch}$ explosion models.
		They are compatible with models assuming either single- or double-degenerate progenitor configurations. We also recover a sub-sample of supernovae within $1.2 $ M$_\odot$ $< M_{\rm {ej}} < 1.5$ M$_\odot$ that are consistent with near-Chandrasekhar explosions. Taking into account the uncertainties of the inferred ejecta masses, about half of our SNe are compatible with both explosion models. We compare our results with those in previous studies, and discuss the caveats and concerns regarding the applied methodology. 
	\end{abstract}


	\section{Introduction} \label{sec:intro}
	
	Both the progenitor configuration and the explosion mechanism of thermonuclear supernovae (Type Ia SNe) are still heavily debated, despite the intense and extensive studies that such events have received for $50+$ years \citep[see. e.g.,][for reviews]{maoz14, bw17, lm18}. It has been recognized decades ago that SNe Ia are produced by thermonuclear explosions of carbon-oxygen white dwarfs (C/O WDs). It is also known that the WD must be in a binary system to be able to reach the physical state that leads to explosion. There are, however, many possible configurations for such a binary system \citep[see, e.g., Fig. 2 in][]{liu23}. They are usually categorized by the number of WDs in the system: a single-degenerate (SD) configuration contains only one WD, while double-degenerate (DD) systems consist of two WDs \citep{lm18}. In addition, the core-degenerate scenario was also proposed \citep{lr03}, in which the WD and the degenerate core of an AGB star rapidly merge during a common-envelope phase. Merging also plays a key role in the DD scenario, although an interesting possibility of a direct WD--WD collision induced by a third body was also considered as an alternative scenario \citep{thompson11}. A detailed review of the pros and cons of all these possible progenitor configurations can be found in, e.g., \citet{lm18}. For further discussion, including a detailed list of references, we refer to \citet{liu23}. 
	
	There are also several possibilities for the explosion mechanism. The key parameter is the mass of the exploding WD, i.e., whether it is close to the Chandrasekhar mass or significantly below it (near-Chandra, or sub-Chandra models, respectively). \citet{liu23} gives a detailed summary of the relevant explosion mechanisms, see their Fig.~3 and Table~2. 
	
	The high degree of homogeneity of the observables of SNe Ia, including light curves (LCs) and spectra, 
	can be explained relatively easily by having $\sim 0.5$ M$_\odot$ of $^{56}$Ni surrounded by a layer consisting of mostly $^{28}$Si, $^{12}$C and $^{16}$O and expanding homologously with $v_{\rm max} \approx 20,000$ km~s$^{-1}$ \citep[e.g.,][]{hoeflich17}. Such an ejecta configuration can be produced by a variety of progenitor systems and explosion mechanisms, involving WDs having either near-Chandra or sub-Chandra masses. Historically, 
	SNe Ia have been thought to arise from the explosion of near-Chandra WDs for many decades. 
	More recent studies found, however, that ejecta masses derived directly from photometric observations \citep{scalzo14a, scalzo14b, scalzo19, ktr20} resulted in a significant fraction of sub-Chandra masses, which actually turned out to be more numerous than near-Chandra events among the normal SNe Ia (excluding 91T-like and super-Chandra SNe). 
	
	Nebular spectroscopy is another powerful tool for constraining ejecta masses. Recent analyses of SNe Ia nebular spectra, both in the optical and near-infrared, compared the observed strength of (forbidden) emission lines of iron peak elements  (mostly Ni and Fe) with the theoretical predictions to reveal the possible progenitor systems and explosion types. These studies suggested a mixed population of sub-Chandra and near-Chandra progenitors for the observed events. For example, \citet{maguire18}, \cite{diamond18} and \citet{dhawan18} found consistency with near-Chandra delayed-detonation models, while \citet{flors18, flors20} preferred mostly sub-Chandra explosions. Most recently, \citet{liu23_2} found that $\sim$ 67\% of their sample (24 out of 36 SNe) is consistent with sub-Chandra explosion models. 
	
	On the other hand, to explain the ionization balance in late-time SN Ia spectra, the presence of neutron-rich Fe-group elements (like $^{55}$Mn and $^{57}$Fe, $^{58}$Ni) is needed \citep[e.g.,][]{lm18}. The production of such species and their corresponding decay chains ($^{57}$Ni$\rightarrow$$^{57}$Co$\rightarrow$$^{57}$Fe, or $^{55}$Co$\rightarrow$$^{55}$Fe$\rightarrow$$^{55}$Mn) require
	high central densities, i.e., near-Chandra masses for the exploding WDs \citep[e.g.,][]{seitan13}. However, this does not rule out sub-Chandra explosions, because those are also capable of producing (at least some fraction of) such neutron-rich isotopes \citep[see, e.g.,][]{shen18}. 
	
	There are also other theoretical arguments that favor near-Chandra, or even super-Chandra WDs at the moment of explosion. In the SD scenario the accreting WD must gain angular momentum as well, resulting in a rapidly spinning, highly magnetized object. Such a WD may significantly exceed the Chandrasekhar mass due to centrifugal forces and magnetic pressure. It must lose its angular momentum in order to reach central density of $\sim 3 \times 10^9$ g~cm$^{-3}$ to explode. Such spin-up/spin-down models might explain the existence of Ia-like super-Chandra explosions, but it is uncertain what fraction of the near-Chandra events they represent in reality \citep[see][and references therein]{bw17}.
	Another possibility is the violent merger scenario, arising from double degenerate systems, where the merger of two sub-Chandra C/O WDs could lead to a super-Chandra explosion \citep[e.g.,][]{pakmor12, pakmor13}.
	
	Motivated by the observational findings mentioned above, a considerable number of Ia explosion models that involve sub-Chandra ($0.9 \lesssim M \lesssim 1.2$~M$_\odot$, \citep{blondin17, shen18, polin19}) WDs have been published recently \citep[see, e.g.,][and references therein]{liu23}. Their common feature is the capability of modeling the ignition of the thermonuclear explosion self-consistently, without the need for an artificial triggering or a by-hand implementation of the deflagration-to-detonation transition, either via double-detonation (DuDe, i.e., triggering the explosion of the C/O WD by the detonation of a $\sim 0.01 ~ M_\odot$ He layer on top of the WD), or by the ``dynamically-driven double-degenerate double-detonation'' (D$^6$, i.e., He-detonation caused by the merging of a C/O and a He-C/O WD) mechanism. Confronting the predictions of these models with the observations may advance our understanding about the true nature of these exotic explosions.
	
	The goal of the present paper is the continuation of the work by \citet{ktr20}, hereafter referred as KTR20, by extending the photometric sample of SNe Ia with more well-observed objects, to derive constraints on the physical properties of SNe Ia, especially the ejecta mass, by combining photometric and spectroscopic data. In Section~2 we describe the new observational data used in this paper. In Section~3 we present the applied methods for distance determination, while Section~4 contains the details about the construction of the bolometric light curves. The method of inferring the physical parameters from the observations is described in Section~5. The results are presented and discussed in Section~6, while in Section~7 we summarize our conclusions. 
	
	\section{Data} \label{sec:data}
	
	In this section we present the observational data, the instruments utilized for the present study and the applied data reduction methods.
	
	\subsection{Observations}   \label{sec:obs}
	
	We carried out photometric observations of 28 Type Ia SNe between 2019 and 2023 from the Piszk\'estet\H{o} mountain station of Konkoly Observatory, Hungary in order to continue the photometric study of SNe Ia presented in KTR20. Our selection criteria for the sample SNe were the following: 
	\begin{itemize}
		\item Declination above $-15$ degrees, in order to be observable from Piszk\'estet\H{o} station. 
		\item Low redshift, i.e., $0 < z < 0.1$. The minimum redshift in the sample is  $z = 0.003$, while the maximum is $z = 0.07$.
		\item Pre-maximum discovery date in order to get reliable estimates on the rise time of the LCs.
		\item Follow-up observations up to at least 40 days after maximum light.
	\end{itemize}
	
	From the overall sample of 28 objects, 21 exploded in spiral galaxies, 3 in lenticular galaxies, 3 in ellipticals and 1 host has unknown morphology. 
	
	The data were taken using the 0.8m Ritchy--Chrétien (RC80) telescope of Konkoly Observatory, manufactured by the AstroSysteme Austria. RC80 is equipped with a 2048 $\times$ 2048 back-illuminated FLI PL230 CCD chip that has 0.55" pixel scale and Johnson--Cousins {\it BV} and Sloan {\it griz} filters. This instrument is appropriate to obtain photometry for nearby ($z < 0.1$) Type Ia SNe with acceptable photometric signal-to-noise (S/N) of $\gtrsim 10$. 
	
	There is an identical telescope, called BRC80, located at Baja Observatory of the University of Szeged, Hungary. Photometric data for two objects in the sample (SN~2021afsj and SN~2022fw) were obtained with this instrument. In the case of two other SNe (SN~2019bkh and SN~2019np) we also used some frames taken with the 0.6/0.9m Schmidt telescope of the Piszkéstető Observatory using Johnson--Cousins $BVRI$ filters, in order to extend the pre-maximum coverage of their LCs. The technical details of the Schmidt telescope can be found in KTR20, Section~2.
	
\begin{figure*}
		\centering
		\includegraphics[width=13cm]{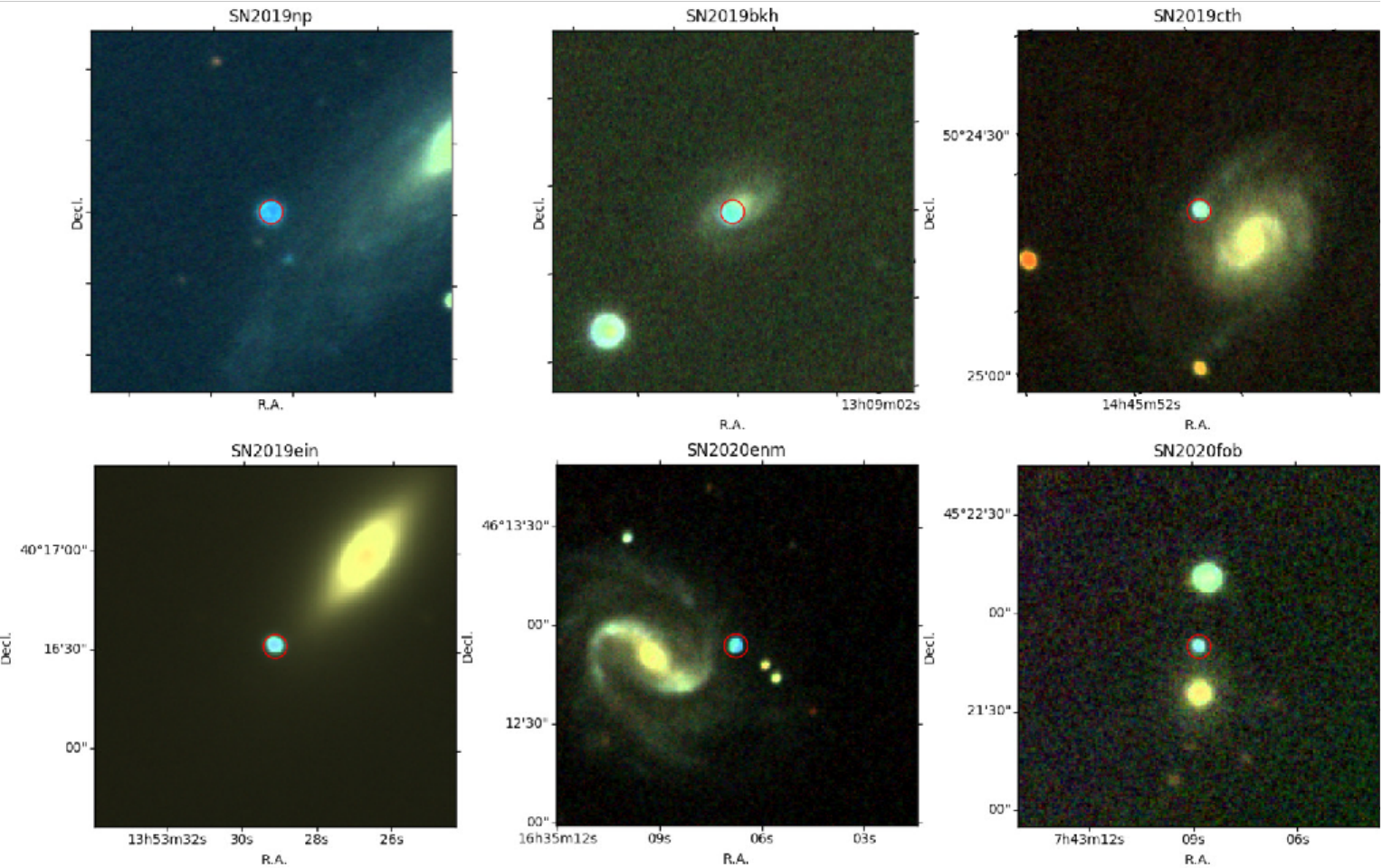}
		\includegraphics[width=13cm]{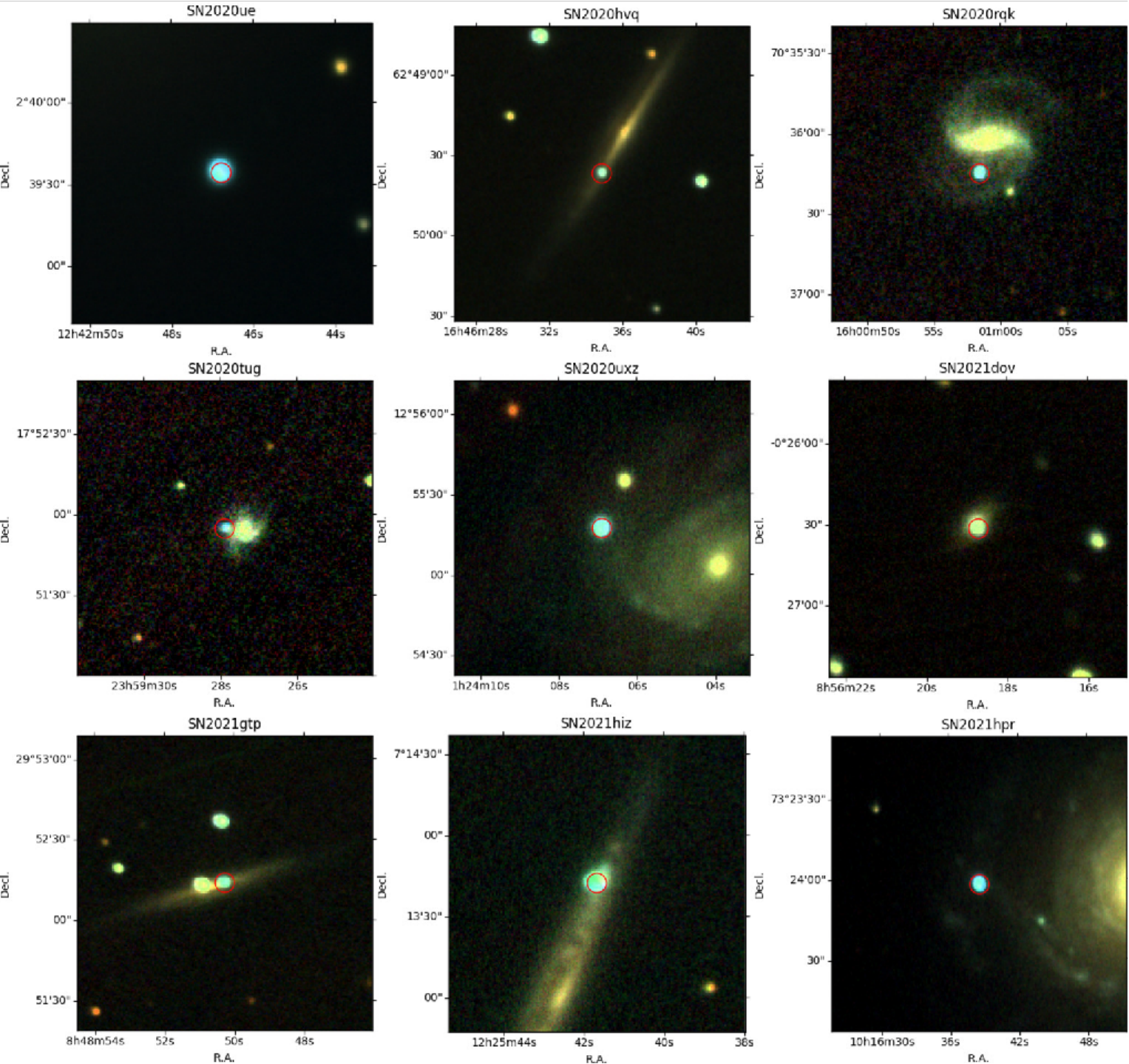}
		\caption{Color-combined $gri$ images centered on the studied SNe. The size of each frame is $1.53 \times 1.53$ arcmin$^2$.}
		\label{fig:stamps1}
	\end{figure*}

\begin{figure*}
		\centering
		\includegraphics[width=13cm]{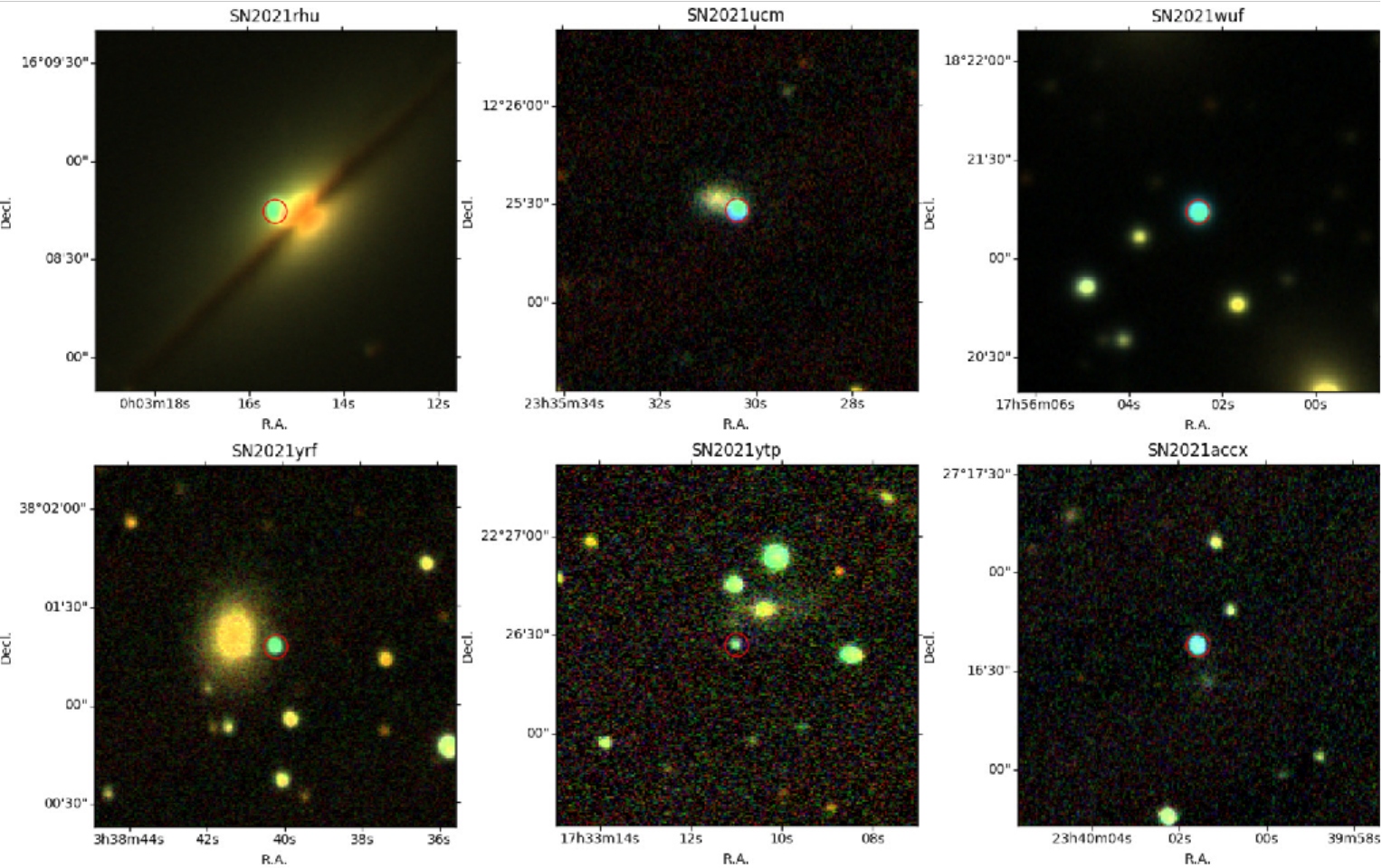}
		\includegraphics[width=13cm]{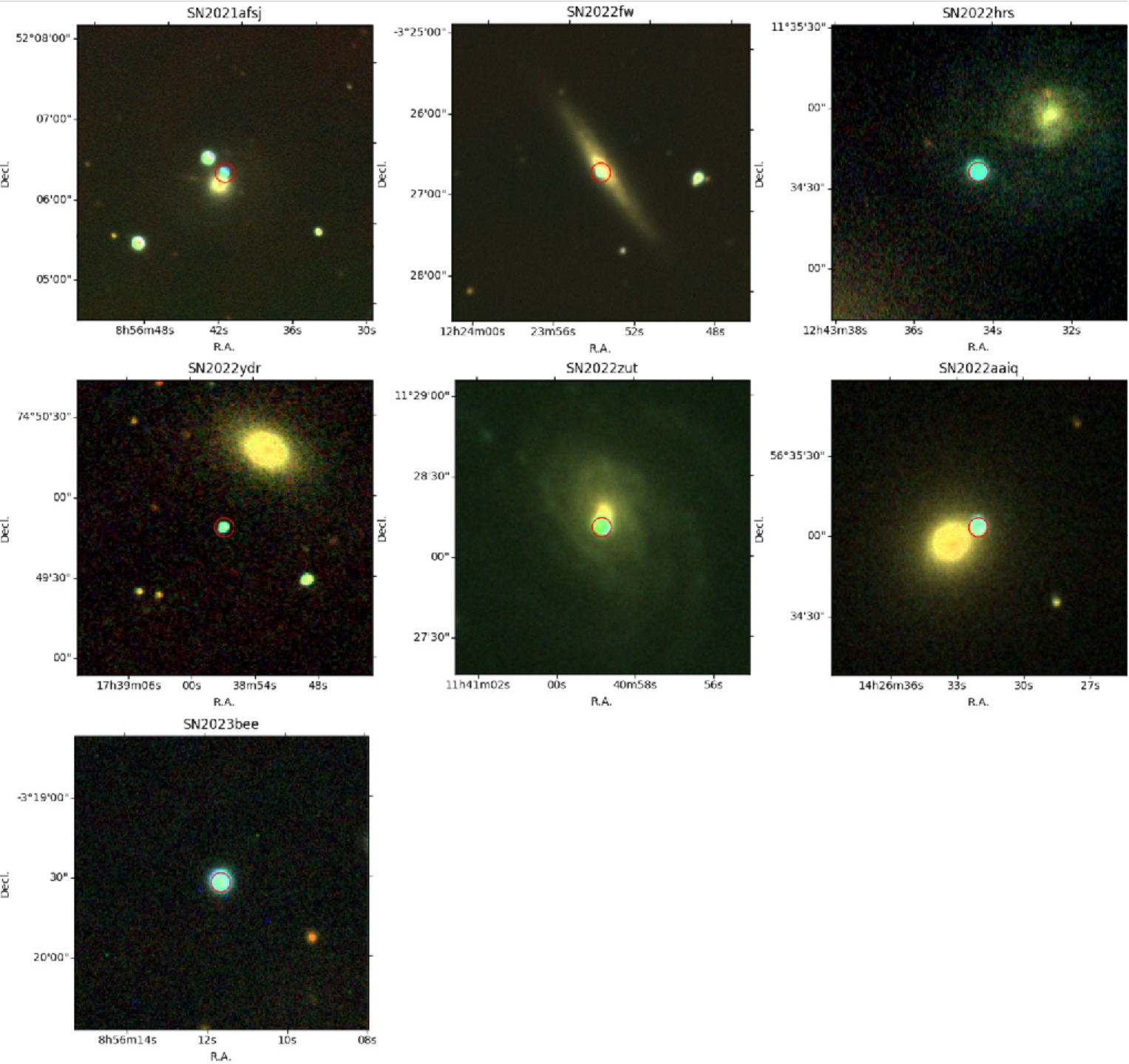}
		\caption{The same as Figure~\ref{fig:stamps1} but for additional SNe in the sample.}
		\label{fig:stamps2}
	\end{figure*}

	The basic data of the studied SNe Ia are collected in Table~\ref{tab:basic}. The total (Milky Way plus host) reddenings ($E(B-V)_{tot}$) and the luminosity distances ($D_L$) were derived by fitting the SN LCs with MLCS2k2 and SALT3 (see Section~\ref{sec:distances} for details). 
	The color-combined $gri$ images of the studied SNe are displayed in Figures~\ref{fig:stamps1} and \ref{fig:stamps2}.
	
	\begin{table*} 
		\tiny
		\caption{Basic data of the studied SNe Ia.}
		\begin{tabular}{lcccccccccc}
			\hline
			SN  & R.A. & Dec. & $E(B-V)$  & $E(B-V)$ & Discovery & Host galaxy  &  z  &  Type & $D_{\rm {L}}$ \\
			&  (hh:mm:ss) & (dd:mm:ss) & (MW, mag) & (total, mag) &  &              &     &  & (Mpc)    \\
			\hline
			SN~2019bkh  & 13:08:57.4 & +28:16:52.4 & 0.009 & 0.119 & 2019-03-03  & MCG +05-31-141  &  0.0195 & Ia-norm & 113.2 (2.2)\\
			SN~2019ein  & 13:53:29.1 & +40:16:31.3 & 0.010 & 0.020 & 2019-05-01 & NGC 5353  &  0.0078 & Ia-norm &  42.2 (0.8)  \\
			SN~2019cth   & 14:45:47.5 & +50:23:48.3  & 0.024 & 0.063 & 2019-04-03   & IC 1056   & 0.0134  & Ia-91T & 63.4 (1.3) \\ 
			SN~2019np   & 10:29:22.0 & +29:30:38.3 & 0.017 & 0.088 & 2019-01-09   & NGC 3254  &  0.0044   & Ia-norm & 32.6 (0.6) \\
			SN~2020enm   & 16:35:06.8 & +46:12:53.3  & 0.015 & 0.137 & 2020-03-16    & IC 1222   &   0.0308 & Ia-91T & 141.3 (2.7) \\
			SN~2020fob  & 07:43:08.8 & +45:21:49.4 & 0.042 & 0.055 & 2020-03-30   & CGCG 235-032  & 0.0315   & Ia-norm  &  150.7 (3.5)  \\
			SN~2020hvq   & 16:46:34.9 & +62:49:36.8 & 0.036 & 0.077 & 2020-04-21  & UGC 10561  & 0.0190  & Ia-norm &  97.0 (3.6) \\
			SN~2020rqk  & 16:00:58.5 & +70:36:15.0  & 0.026 & 0.058 & 2020-08-19  & NGC 6071  &  0.0239  & Ia-norm & 117.4 (2.2) \\
			SN~2020tug   & 23:59:27.9 & +17:51:54.7 & 0.025 & 0.080 & 2020-09-19  & KUG 2356+175  &  0.0500   & Ia-norm &  247.5 (7.6) \\
			SN~2020ue   & 12:42:46.8 & +02:39:34.1 & 0.024 & 0.024 & 2020-01-12   &  NGC 4636 &  0.0030  & Ia-norm  & 16.4 (0.3) \\
			SN~2020uxz  & 01:24:06.9 & +12:55:17.2  & 0.033 & 0.033 & 2020-10-05  &  NGC 514 & 0.0083   & Ia-norm  & 38.7 (0.7) \\
			SN~2021accx  & 23:40:01.6 & +27:16:37.4  & 0.083 & 0.112 & 2021-10-23  & J234001.30+271626.4 \tablenotemark{a} & 0.0310   & Ia-norm & 144.8 (2.5)  \\
			SN~2021afsj & 08:56:42.4 & +52:06:28.8 & 0.017 & 0.124 & 2021-11-29 & UGC 4671 & 0.0122 & Ia-norm  &  60.8 (1.4)  \\
			SN~2021dov   & 08:56:18.7 & -00:26:31.7 & 0.030 & 0.282 & 2021-02-22   & Z 5-38  &   0.0127 & Ia-91T &  58.7 (1.0)  \\
			SN~2021gtp & 08:48:50.3 & +29:52:13.8 & 0.035 & 0.235 & 2021-03-19 & UGC 04611 & 0.0197 & Ia-norm  &  94.2 (3.5) \\
			SN~2021hiz  & 12:25:41.7 & +07:13:42.2 & 0.022 & 0.116 & 2021-03-30   & UGC 7513  & 0.0033   & Ia-norm  & 27.9 (0.5) \\
			SN~2021hpr  & 10:16:38.6 & +73:24:01.8  & 0.022 & 0.109 & 2021-04-02  &  NGC 3147 &  0.0094  & Ia-norm  &  46.8 (1.2)  \\
			SN~2021rhu   & 00:03:15.4 & +16:08:44.5 & 0.038 & 0.129 & 2021-07-01  & NGC 7814  & 0.0035  & Ia-norm  &  16.6 (0.3)  \\
			SN~2021ucm  & 23:35:30.4 & +12:25:27.9 & 0.059 & 0.166 & 2021-07-25  &  J233530.78+122531.4 \tablenotemark{a} &  0.0240 & Ia-norm   &  113.7 (2.3)  \\
			SN~2021wuf   & 17:56:02.5 & +18:21:14.1 & 0.078 & 0.091 & 2021-08-23  & NGC 6500  &  0.0100   & Ia-91T & 44.2 (1.0) \\
			SN~2021yrf  & 03:38:40.2 & +38:01:17.7 & 0.329 & 0.387 & 2021-09-09  & CGCG 525-046  &  0.0166   & Ia-norm & 86.1 (1.6)  \\
			SN~2021ytp   & 17:33:11.0 & +22:26:27.1 & 0.049 & 0.153 & 2021-09-09   & J173310.37+222637.5 \tablenotemark{a}  & 0.0650  & Ia-norm   &  279.7 (4.5) \\
			SN~2022aaiq & 14:26:32.0 & +56:35:03.1 & 0.017 & 0.146 & 2022-11-15 & NGC 5631 & 0.0065 & Ia-norm  &  35.8 (1.9)  \\
			SN~2022fw   & 12:23:54.0 & -03:26:37.9 & 0.032 & 0.290 & 2022-01-09 & NGC 4348   & 0.0067 & Ia-norm  & 31.4 (1.2)  \\
			SN~2022hrs & 12:43:34.3 & +11:34:35.8 & 0.023 & 0.188 & 2022-04-16 & NGC 4647 & 0.0047 & Ia-norm  & 19.1 (0.3)  \\
			SN~2022ydr & 17:38:56.8 & +74:49:48.5 & 0.032 & 0.032 & 2022-10-18 & UGC 10949 & 0.0249 & Ia-91bg  & 133.8 (5.5) \\
			SN~2022zut & 11:40:58.8 & +11:28:10.9 & 0.038 & 0.325 & 2022-11-09 & NGC 3810 & 0.0033 & Ia-norm  & 18.8 (0.4)  \\
			SN~2023bee & 08:56:11.6 & -03:19:32.0 & 0.014 & 0.014 & 2023-02-01 & NGC 2708 & 0.0067 & Ia-norm  &  37.5 (1.1) \\
			
			\hline         
		\end{tabular}
		\tablenotetext{a}{Host galaxy names in the SDSS catalogue.}
		
		\label{tab:basic}
	\end{table*}
	\normalsize

	\subsection{Reduction \& photometry}   \label{sec:phot}

	The basic data reduction steps (bias, dark and flatfield corrections) were done in the standard way using the appropriate tasks in IRAF\footnote{IRAF is distributed by the National Optical Astronomy Observatories, which are operated by the Association of Universities for Research in Astronomy, Inc., under cooperative agreement with the National Science Foundation. http://iraf.noirlab.edu} (Image Reduction and Analysis Facility). The frames were then registered and median-combined to produce a single frame in each filter \citep[see, e.g.,][]{vinko18}. 
	
	Except in a few cases, when the supernova occured so far away from its host galaxy that it was unaffected by the host galaxy background, we applied image subtraction photometry to separate the light of the host galaxy from the light of the transient. Since we do not have images with the RC80 telescope taken before or sufficiently late after explosion, we used frames from the PanSTARRS-1 (PS1) imaging archive \citep{flewelling20} for the image subtraction process. Before subtracting the PS1 images, we applied the IRAF \texttt{geomap}, \texttt{gregister}, \texttt{psfmatch} and \texttt{linmatch} tasks to crop, rotate, and scale the PS1 images to match the reduced RC80 frames. After removing the host from the frames, we determined the apparent magnitude of the SN and multiple local comparison stars in each filter via aperture photometry. We used the catalogued PS1 magnitudes of the comparison stars for transforming the instrumental magnitudes to the standard system in the standard way, including color terms in the transformation. For the $B$ and $V$ data the $g_{PS1}$ and $r_{PS1}$ magnitudes of the comparison stars were transformed into Johnson--Cousins $B$ and $V$ magnitudes based on the formulae given by \citet{tonry12}. 
	
	Finally, the uncertainties of the obtained magnitudes were calculated as the squared sum of the photometric errors given by IRAF, and the fitting errors from the standard transformation.
	
	All photometric data can be found online in the following GitHub repository: 
	\url{https://github.com/borazso/PiszkeSN}.
	
	\subsection{Spectroscopy}\label{sec:spectra}
	
	In order to determine expansion velocities (see Section~\ref{sec:velocity}), we used spectra obtained with the instruments at Las Cumbres Observatory (LCO) and McDonald Observatory (Table~\ref{tab:spectra_list}).
	
	The LCO spectra were taken with the FLOYDS spectrographs mounted on the LCO 2m telescopes, via the Global Supernova Project (GSP). The spectra were reduced by a custom pipeline using IRAF tasks \citep{valenti14}. Additional data were taken with the Low Resolution Spectrograph 2 (LRS2) on the 10m Hobby-Eberly Telescope (HET) at McDonald Observatory \citep{lrs2}. Since  
	the Si~II $\lambda$6355 line, which was used for the velocity measurements, falls within the wavelength coverage of the blue arm (LRS2-B, 3640–6970 \AA), we used spectra taken with LRS2-B. The spectra were extracted from the IFU data cubes by the {\tt Panacea}\footnote{{\tt https://github.com/grzeimann/Panacea}} pipeline. The extracted spectra were then corrected for redshift and telluric lines using Python scripts and IRAF tasks.

	\subsection{Public databases}   \label{sec:public}
	
	We collected additional data to supplement our observations from two sources. We used public photometry from the Zwicky Transient Facility (ZTF) \citep{bellm19} for 7 SNe (SN~2019bkh, SN~2019ein, SN~2020fob, SN~2020hvq, SN~2020tug, SN~2020uxz, SN~2021ytp), where the RC80 LCs did not cover the pre-maximum phases. Since our goal is to determine the physical parameters of these supernovae, wide phase coverage of the LCs is crucial. 
	Additionally, some physical parameters of the explosion can only be determined by examining the late-time tail of the light-curve. Therefore in the case of 3 SNe (SN~2019ein, SN~2020tug, SN~2020uxz), we used the public ZTF photometry to extend their LCs into later phases. The downloaded ZTF data can be found in the GitHub repository\footnote{\tt https://github.com/borazso/PiszkeSN} of this paper. 
	We also collected all available public spectra for our sample SNe from the Transient Name Server (TNS)\footnote{\tt https://www.wis-tns.org/}. All spectroscopic data are shown in~Table~\ref{tab:spectra_list}. Details on determining the expansion velocities from these spectra are given in Section~\ref{sec:velocity}.
	
	\begin{table*}
		\tiny
		\centering
		\begin{tabular}{lcccr}
			\hline
			SN  & Date & Phase  & $v_{\rm {exp.}}$ &    Source \\
			& (YYYY-MM-DD)     & (days) & (km\,s$^{-1}$)              &   \\
			\hline
			SN~2019bkh  &   2019-03-04  & -10.75    & 13753 (17) & TNS: ZTF \citep{ztf19bkh} \\
			& 2019-03-05    &    -9.77  &  13362 (749) & TNS: Padova-Asiago \citep{padovaasiago19bkh} \\
			SN~2019cth  & 2019-04-24    & 0.32      & 13577 (81) &  LCO GSP\\
			SN~2019ein  &  2019-05-02   & -14.65     &   22794 (959)  &   TNS: Global SN Project \citep{global19ein} \\
			&  2019-05-05   & -11.68     & 19075 (817)   &  TNS: ZTF  \citep{ztf19ein} \\
			& 2019-05-12    &  -4.73   & 14436 (559) &  TNS: Pascal Le Dû  \citep{du19ein} \\
			& 2019-05-14    &  -2.75   & 13759 (285) &  TNS: Pascal Le Dû \citep{du19ein} \\
			&  2019-05-16   & -0.76     & 12874 (433) &   LCO GSP \\
			& 2019-05-21    &  +4.20     & 12156 (446)  &  TNS: Pascal Le Dû \citep{du19ein}  \\
			SN~2019np   & 2019-01-10    &   -17.35  & 14672 (156)  &  TNS: Global SN Project \citep{global19np} \\
			& 2019-01-10    &  -16.66   & 14973 (400) &   TNS: YNAO \citep{ynao19np} \\
			& 2019-01-12    &  -15.36   &  13623  (759)  &  TNS: ZTF \citep{ztf19np} \\
			& 2019-01-24    &  -3.41    & 10030 (85)   &   LCO GSP \\
			SN~2020enm  & 2020-03-18    &  -11.83   & 11438 (937) &   TNS: YNAO \citep{ynao20enm} \\
			SN~2020fob  & 2020-03-31    &  -7.76    & 11545 (933)  &  TNS: ZTF \citep{ztf20fob} \\
			SN~2020hvq  & 2020-04-22    &  -11.42   & 17033 (587) &   TNS: ISSP \citep{balcon20hvq} \\
			& 2020-04-23    &  -10.40   & 17129 (1303) &  TNS: ZTF  \citep{ztf20hvq} \\
			SN~2020rqk  & 2020-08-16    &  -15.66   & 13975 (887) &   TNS: ZTF  \citep{ztf20rqk} \\
			SN~2020tug  & 2020-09-22    &  -13.38   & 15007 (589) &   TNS: adH0cc \citep{adhoc20tug} \\
			& 2020-10-05    &  -1.00    & 9612 (1139) &   LCO GSP \\
			SN~2020ue   & 2020-01-28    &  2.20      & 10875 (588)   &  LCO GSP \\
			SN~2020uxz  & 2020-10-05    &  -17.22   & 15160 (282) &  TNS: Global SN Project   \citep{global20uxz} \\
			& 2020-10-20    &  -2.35    & 10493 (97)  &   LCO GSP \\
			SN~2021accx &  2021-10-28   & -8.16     & 10816 (748)  &  TNS: ePESSTO+ \citep{epessto21accx} \\
			& 2021-11-06    &  0.57     & 9887 (283)  &   LCO GSP \\
			& 2021-11-11    &  +5.42    & 9907 (283)   &   HET  \\
			SN~2021afsj & 2021-11-30    &  -13.18   & 12681 (1502)  &  TNS: ISSP \citep{balcon21afsj} \\
			SN~2021dov  & 2021-03-12    &  +0.19    & 10478 (388)  &   LCO GSP \\
			& 2021-03-21    &  +9.08    & 8916 (184)  &   TNS: PSH  \citep{psh21dov} \\
			SN~2021gtp  & 2021-03-22    &  -13.75   & 16957 (1654) &  TNS: ZTF \citep{ztf21gtp} \\
			SN~2021hiz  & 2021-03-31    &  -16.62   & 14607 (736) &   TNS: SUCSC \citep{sucsc21hiz} \\
			& 2021-04-17    &  +0.32    & 9787 (309)  &   LCO GSP \\
			SN~2021hpr  & 2021-04-03    &  -15.88   & 19246 (678) &   TNS: Three Hills Observatory \citep{leadbeater21hpr} \\
			& 2021-04-04    &  -14.89   & 18933 (972) &   TNS: ZTF  \citep{ztf21hpr} \\
			& 2021-04-19    &  -0.03    & 11136 (29) &   TNS: PSH \citep{psh21hpr} \\
			SN~2021rhu  & 2021-07-01    &  -14.94   & 17158 (126)  &  TNS: ALeRCE  \citep{alerce21rhu} \\
			& 2021-07-05    &  -10.96   & 13036 (1038) &  TNS: ZTF  \citep{ztf21rhu}  \\
			& 2021-07-14    &   -1.99   &   11600 (263) &   LCO GSP \\
			& 2021-07-22    &  +5.98    & 10479 (489)  &  HET     \\
			& 2021-07-28    &  +11.96   & 9806 (1577) &   HET     \\
			SN~2021ucm  & 2021-08-02    &  -6.88    & 13374 (161) &  TNS: ePESSTO+ \citep{epessto21ucm} \\
			& 2021-08-02    &  -6.34    & 13091 (1277) &  TNS: SCAT  \citep{scat21ucm} \\
			SN~2021wuf  & 2021-08-25    &  -10.65   & 19400 (4546) &  TNS: SCAT \citep{scat21wuf} \\
			& 2021-09-06    &    +1.23  & 12270 (511) &   LCO GSP\\
			SN~2021yrf  & 2021-09-13    &  -11.65   & 16274 (1771) &  TNS: ATLAS \citep{atlas21yrf} \\
			SN~2021ytp  & 2021-09-15    &  -6.10     & 16489 (328)  &  TNS: ePESSTO+  \citep{epessto21ytp} \\
			SN~2022aaiq & 2022-11-17    &  -12.26   & 11253 (349) &   TNS: LICK-UCSC  \citep{ucsc22aaiq}  \\
			& 2022-12-12    &  +12.58   & 9391 (503)  &   TNS: ISSP \citep{balcon22aaiq} \\
			SN~2022fw   & 2022-01-10    &  -14.76   & 13240 (126) &  TNS: Global SN Project  \citep{global22fw} \\
			& 2022-01-11    &  -13.76   & 12853 (583) &  TNS: ePESSTO \citep{epessto22fw} \\
			SN~2022hrs  & 2022-04-16    &  -15.07   & 16821 (2122) &  TNS: ISSP  \citep{balcon22hrs}  \\  
			& 2022-04-21    &  -10.10   & 14353 (705) &   TNS: PSH \citep{psh22hrs} \\
			& 2022-04-29    &  -2.13    & 12587 (943)  &   LCO GSP \\
			SN~2022ydr  & 2022-10-25    &  -5.94    & 10159 (510)  &  TNS: NOT/ALFOSC \citep{alfosc22ydr}  \\
			& 2022-10-27    &  -3.99    & 10688 (788)  &  TNS: ISSP \citep{balcon22ydr} \\
			SN~2022zut  & 2022-11-10    &  -12.31   & 17423 (797) &  TNS: SGLF/LTbot \citep{sglf22zut} \\
			& 2022-11-16    &  -6.33    & 13249 (951)  & HET \\
			& 2022-11-17    &  -5.33    & 12997 (45) & HET \\
			& 2022-11-23    &  +0.65    & 10827 (194)  & LCO GSP\\
			SN~2023bee  & 2023-02-01    &  -17.80    & 23658 (1763)  &  TNS: LiONS  \citep{lions23bee}    \\
			& 2023-02-02    &  -16.81   & 22207 (1941) &  TNS: Global SN Project   \citep{global23bee}  \\
			& 2023-02-11    &  -7.87    & 12150 (40) &   HET    \\
			& 2023-02-16    &  -2.90    & 11461 (45)  &  LCO GSP   \\
			& 2023-02-17    &  -1.91    & 11444 (358)  &  HET   \\
			\hline
		\end{tabular}
		\caption{Spectroscopic data for the sample SNe.}
		\label{tab:spectra_list}
	\end{table*}

	\section{Distance estimates} \label{sec:distances}
	
	In this section, we describe our methods to estimate the distances and reddenings of our sample SNe Ia by applying two different techniques and codes: SALT3 (see Section~\ref{sec:salt3}) and MLCS2k2 (Section~\ref{sec:mlcs2k2}). 
	
	\subsection{SALT3}   \label{sec:salt3}
	
	In the present paper we utilized the SALT3 code, which is the improved version of SALT2 \citep[Spectral Adaptive Lightcurve Template;][]{guy05, guy07, guy10, betoule14}.
	
	In SALT3 the spectral flux of the supernova is expressed as a function of wavelength ($\lambda$) and the rest-frame phase ($p$) as
	\begin{equation}
		F(p, \lambda) = x_0 \cdot \left[ M_0(p, \lambda) + x_1 \cdot M_1(p, \lambda) \right] \cdot \mathrm{exp}(c \cdot \mathrm{CL}(\lambda)),
		\label{eq:salt3}
	\end{equation}
	where $M_0$, $M_1$ and CL are the calibrated template vectors, while the fitting parameters are the flux normalization $x_0$, the stretch parameter $x_1$ and the color parameter $c$. 
	
	Here we adopt the most recent template vectors by \citet{kenworthy21}, which were trained with $\sim 1200$ spectra covering the wavelength range of 2000 and 11000 \AA. This sample is more than an order of magnitude larger compared to the training sets of previous version of the code, thus, SALT3 has lower uncertainties than, e.g., SALT2.4 \citep{taylor23}. In 2022, the SALT3 code was extended further into the near infrared by \citet{pierel22}.

	In the SALT3 framework the luminosity distances to SNe~Ia can be derived in two main steps. First, the three free parameters ($x_0$, $x_1$ and $c$) are found from fitting Equation~(\ref{eq:salt3}) to a particular SN~Ia light curve. Second, their best-fit values are substituted into the Tripp-equation \citep{tripp98} to get the distance modulus $\mu$:
	\begin{equation}
		\mu = m_B + \alpha x_1 - \beta c - M_B^0,
		\label{eq:tripp}
	\end{equation} 
	where $m_B$ is the rest-frame $B$-band peak magnitude of the SN ($m_B = -2.5 \log_{10} (x_0) + m_0$), $\alpha$ and $\beta$ are the nuisance parameters representing the slopes of the stretch-
	luminosity and color–luminosity relations,  while $M_B^0$ is the fiducial absolute magnitude of a SN Ia in the rest-frame $B$-band. Here we adopt $\alpha = 0.133 \pm 0.003$ and $\beta = 2.846 \pm 0.017$ from \citet{pierel22}, and $M_B^0~=~ -19.253$ from \citet{riess22}. We find $m_0 = 10.607$ by matching our best-fit $m_B$ values to those of the Pantheon+SH0ES\footnote{\tt https://github.com/PantheonPlusSH0ES/DataRelease} SNe~Ia sample \citep{riess22}. 
	Note that we ignore second-order terms in Equation~(\ref{eq:tripp}), like the mass-step $\delta \mu_{\rm  host}$ \citep{betoule14, brout19}, or the bias correction $\delta \mu_{\rm  bias}$ \citep{kessler19}, because for our low-z sample their values are smaller than the typical uncertainty of our distance modulus estimates ($\sim 0.1$ mag). 
	
	The fitting was done with the {\tt SNcosmo}\footnote{\tt https://sncosmo.readthedocs.io/en/stable/} python library \citep{barbary16}, adopting Milky Way extinction estimates provided by the NASA/IPAC Infared Science Archive \citep{schlafly11, schlegel98}, and redshift information from the NASA Extragalactic Database (NED) \footnote{The NASA/IPAC Extragalactic Database (NED) is funded by the National Aeronautics and Space Administration and operated by the California Institute of Technology.}. The best-fit values of $t_0$, $x_0$, $x_1$ and $c$ for each supernova can be found in the GitHub repository of this paper (see above). The final luminosity distances (via $\mu$ from Equation~\ref{eq:tripp}) are shown in Table~\ref{tab:basic}.

	\subsection{MLCS2k2}   \label{sec:mlcs2k2}
	
	Even though SALT3 provides accurate relative distances to SNe Ia via Equation~(\ref{eq:tripp}), which are now tied to Cepheid-based absolute distances \citep{riess22}, the construction of the bolometric light curves require information about the extinction due to dust in the host galaxy of each SN. Since the SALT3 flux model (Equation~\ref{eq:salt3}) incorporates the effect of dust extinction into the color parameter, it does not give a direct prediction of the dust extinction occurred in the host. Thus, we utilized the Multi-Color Light Curve Shape Method \citep[MLCS2k2;][]{riess95, riess96, jha07} to overcome this difficulty.

	Since MLCS2k2 was calibrated to Johnson--Cousins $UBVRI$ photometry \citep{jha07} we applied this code only to our Johnson $B$- and $V$-band data (after correcting them to Milky Way extinction, as above), and fixed both the distance modulus and the moment of the $B$-band maximum light from the results of our SALT3 fitting. This way we got reasonable estimates on the host galaxy extinction value $A_V$, from which we inferred the $E(B-V)_\mathrm{host}$ reddening and the extinction in other passbands using the Milky Way reddening law parameter $R_V = 3.1$. Finally, the total reddening was calculated as $E(B-V)_\mathrm{total} = E(B-V)_\mathrm{MW} + E(B-V)_\mathrm{host}$ (see Table~\ref{tab:basic}). 
	The best-fit MLCS2k2 parameters can be found in the GitHub repository\footnote{\tt https://github.com/borazso/PiszkeSN} of this paper.

	\section{Bolometric light curves and their modeling} \label{sec:bolometric}

	In this Section we briefly describe the construction of the bolometric LCs and the fitting of the radiation-diffusion Arnett models. 
	
	\subsection{Bolometric light curve construction} \label{sec:trapez}
	
	The bolometric LCs were constructed with the same methodology as in KTR20: briefly, after correcting the $BVgri$ magnitudes (the $z$-band data were omitted due to their higher uncertainties) to the effect of dust extinction (both in the Milky Way and the host galaxy, see Table~\ref{tab:basic}), the magnitudes were converted to flux densities, then corrected for distance and redshift. The resulting rest-frame absolute flux densities were then integrated along the wavelength axis by applying the trapezoidal rule. The missing UV- and IR-fluxes were approximated in a similar way as in KTR20, by assuming a linear decrease in flux from the wavelength of the $B$ filter down to 1900 \AA, and estimating the IR contribution by fitting a Rayleigh-Jeans tail to the $i$-band flux, then integrating it from the wavelength of the $i$-band filter to infinity. This method was extensively tested by KTR20 concluding that this technique gives a reliable representation of the true bolometric flux/luminosity, and the relative uncertainty of the derived fluxes does not exceed $\sim 10$\%.
	
	As mentioned above, in those cases when our observations did not cover the pre-maximum phases, we extended the light curves with public ZTF photometry. Since ZTF data are available only in $g$- and $r$-bands, applying the trapezoidal integration method leads to higher uncertainties in the resulting pseudo-bolometric fluxes. Thus, we tested the effect of omitting bands other than $g$ and $r$ from the integration using the high-cadence data of SN~2021hpr. We found that restricting the data only to $g$ and $r$ (but doing the correction for the missing bands as above) underestimates the integrated flux by only $\sim 2.8$\% during the pre-maximum phases compared to the case when the full $BVgri$ data are integrated. The difference increases up to $\sim 8$\%
	around maximum light, then drops back to $\sim 2.5$\% later than +5 days from maximum. Since pre-maximum data are important to constrain the physical parameters of the light curve model, motivated by these experiments we decided to use the ZTF-based quasi-bolometric data in the final light curves, but assigned 50\% larger error bars to those points in order to represent their higher uncertainty.

	\subsection{Model fitting}   \label{sec:bol_fit}

	The bolometric LCs were modeled applying the Monte-Carlo based Minim code \citep{manos13}, which fits the radiation-diffusion model of \citet{arnett82} to the quasi-bolometric LC of the SN. The following assumptions were adopted for the Arnett-model: 
	\begin{itemize}
		\item{spherically symmetric, homologously expanding ejecta,}
		\item{spatially constant density profile,}
		\item{constant optical opacity, both spatially and temporally,}
		\item{centrally located radioactive heating source,}
		\item{radiation pressure being dominant,}
		\item{the spatial temperature profile being fixed to the ``radiative-zero'' solution, }
		\item{separation of the spatial and temporal part of the energy equation.}
	\end{itemize}

	The Minim code uses the Price-algorithm in order to localize the absolute minimum of the $\chi^2$ surface in the parameter space within pre-selected parameter bounds. The detailed description of the code and its method to estimate the uncertainties of each fitted parameter values can be found in \citet{manos13}.
	
	The SNe Ia LCs were fitted with the usual model of radioactive decay of $^{56}$Ni~$\rightarrow$~$^{56}$Co~$\rightarrow$~$^{56}$Fe by optimizing the following four parameters (e.g., KTR20): 
	
	\begin{itemize}
		\item $t_0$: the moment of the explosion with respect to the date of  maximum brightness in the $B$-band (in days),
		\item $t_{\rm LC}$: the LC timescale, similar to the rise time to maximum (in days),
		\item  $t_{\gamma}$: the timescale of the gamma-ray leaking (in days),
		\item $M_{\rm Ni}$: the initial nickel-mass (in M$_\odot$).
	\end{itemize}
	
	The best-fit model parameters are given in the first 4 columns of Table~\ref{tab:phys_prop}.
	
	\section{Physical parameters} \label{sec:phys_par}
	
	To determine the physical parameters for each SN, such as the ejecta mass ($M_{\rm ej}$) or the opacity ($\kappa$), we used an improved version of the method presented in \citet{li19} and KTR20 (see Equations~\ref{eq:mej} and \ref{eq:kappa}). In the present paper the expansion velocity $v_{\rm exp}$ was estimated directly from spectra taken near maximum light (described in Section~\ref{sec:velocity}). By using the timescales $t_{\gamma}$ and $t_{\rm LC}$ provided by the Minim code, we calculate the ejecta mass as
	\begin{equation}
		M_{\rm ej} = \frac{4 \pi t_{\gamma}^2 \cdot v_{\rm exp}^2}{3  \kappa_{\gamma}},
		\label{eq:mej}
	\end{equation}
	and the average opacity in the ejecta as
	\begin{equation}
		\kappa = \frac{t_{\rm LC}^2 \cdot \beta c \cdot v_{\rm exp}}{2  M_{\rm ej}},
		\label{eq:kappa}
	\end{equation}
	where $\kappa_{\gamma}~=~ 0.025 \text{ cm}^2\text{g}^{-1}$ is the opacity for gamma-ray photons \citep{guttman24}, $\beta~=~13.8$ is an integration constant related to the density distribution of the ejecta \citep{arnett82}, and $v_{\rm exp}$ is the expansion velocity of the SN.
	
	Furthermore, the kinetic energy of the expanding ejecta (assuming homologous expansion of a constant density sphere) can be expressed as
	\begin{equation}
		E_{\rm kin} = \frac{3}{10}  M_{\rm ej} \cdot v_{\rm exp}^2.
		\label{eq:ekin}
	\end{equation}

	\subsection{Expansion velocities}   \label{sec:velocity}

	\begin{figure}[H]
		\centering
		\includegraphics[scale = 0.6]{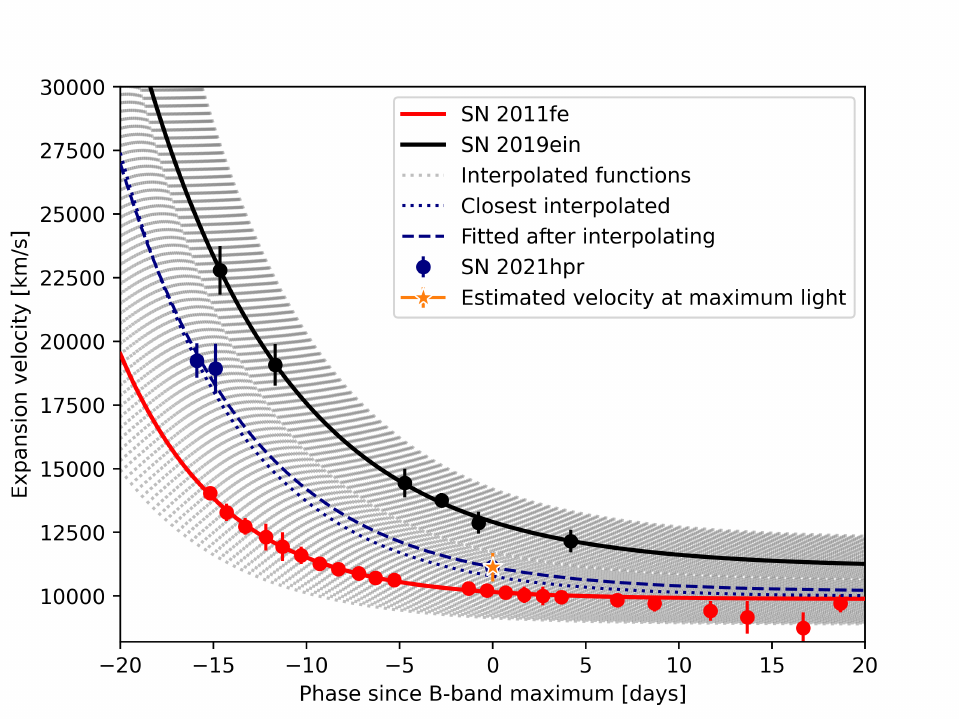}
		\caption{Visualization of the method we used to estimate $v_{exp}$ for SN~2021hpr. The high- and low velocity bounds are defined by SN~2019ein (black symbols and thick curve) and SN~2011fe (red symbols and thick curve). Among the interpolated velocity curves (thin dotted curves) the best-fit curve for SN~2021hpr (blue dashed curve) is used to estimate the velocity at maximum (orange symbol).}
		\label{fig:interpol}
	\end{figure}
	
	The expansion velocity, $v_{\rm exp}$, appearing in Equations (\ref{eq:mej}), (\ref{eq:kappa}) and (\ref{eq:ekin}), was estimated from optical spectra taken near maximum light (see Section~\ref{sec:spectra}): we used the Doppler-shift of the absorption minimum of the Si~II $\lambda$6355 feature in each available spectrum to measure the expansion velocity at the given phase. After correcting the spectra for the redshift of the host galaxy, the observed wavelength of the Si~II feature was measured by fitting the absorption component with a single Gaussian function near the core, then taking the center of the fitted profile as the observed wavelength ($\lambda_{obs}$). The Doppler-shift of $\lambda_{obs}$ from the rest-frame position, $\lambda_0 = 6355$ \AA, gives the Si~II velocity ($v_{\rm SiII}$). In this paper we adopted $v_{\rm SiII}$ measured at maximum light as an estimate for $v_{\rm exp}$. 
	
	For those SNe that had available spectra taken close to maximum, the determination of $v_{\rm exp}$ was relatively easy. However, this was not the case for most of our sample SNe.
	In these cases we had to interpolate between the $v_{\rm SiII}$ values measured at different phases in order to get the velocity near maximum light. The interpolation was done using a pre-defined set of exponential functions fitted to the measured $v_{\rm SiII}$ velocities of the well-observed SN~2019ein (a high-velocity SN Ia) and SN~2011fe (a normal velocity SN Ia). 
	
	For SN~2019ein we used the spectra listed in Table~\ref{tab:spectra_list}, while for SN~2011fe we collected the spectra from \citet{pereira13}. First, the measured velocities of these two SNe as a function of time were fitted with exponential functions. Then, 100 other exponentials were computed to model the temporal evolution of the velocity within the range defined by SN~2011fe and SN~2019ein (plotted with grey dotted curves in Figure~\ref{fig:interpol}). The parameters of these exponentials were computed starting from 90\% of the parameters of SN~2011fe, extending up to 110\% of the parameters of SN~2019ein. 
	
	To estimate the velocity at the moment of maximum light for a given SN, first, we plotted the measured velocities against phase, then looked for the best-fitting pre-defined exponential that was closest to the measured velocities. Then, using the parameters of this exponential we inferred its value at zero phase, i.e., at the moment of maximum light in the B-band (see the blue dashed curve in Figure~\ref{fig:interpol} as an example for SN~2021hpr). 
	
	This method worked well in those cases when only 1 or 2 spectra were available. However, for SNe with multiple spectra available, we found that some of them showed a velocity evolution that was slightly different from the pre-defined exponential functions. Thus, in those cases we preferred fitting the measured velocities with another exponential function and using this best-fit exponential to estimate the velocity at maximum light. 
	This fitting introduces a small ($\sim 200$ km\,s$^{-1}$) error and so the uncertainty of the Si II velocity at maximum light comes in most part from the uncertainty of how well we can find the line minima in the individual spectra.
	
\begin{figure*}[ht]
		\centering
		\includegraphics[width=0.8\textwidth]{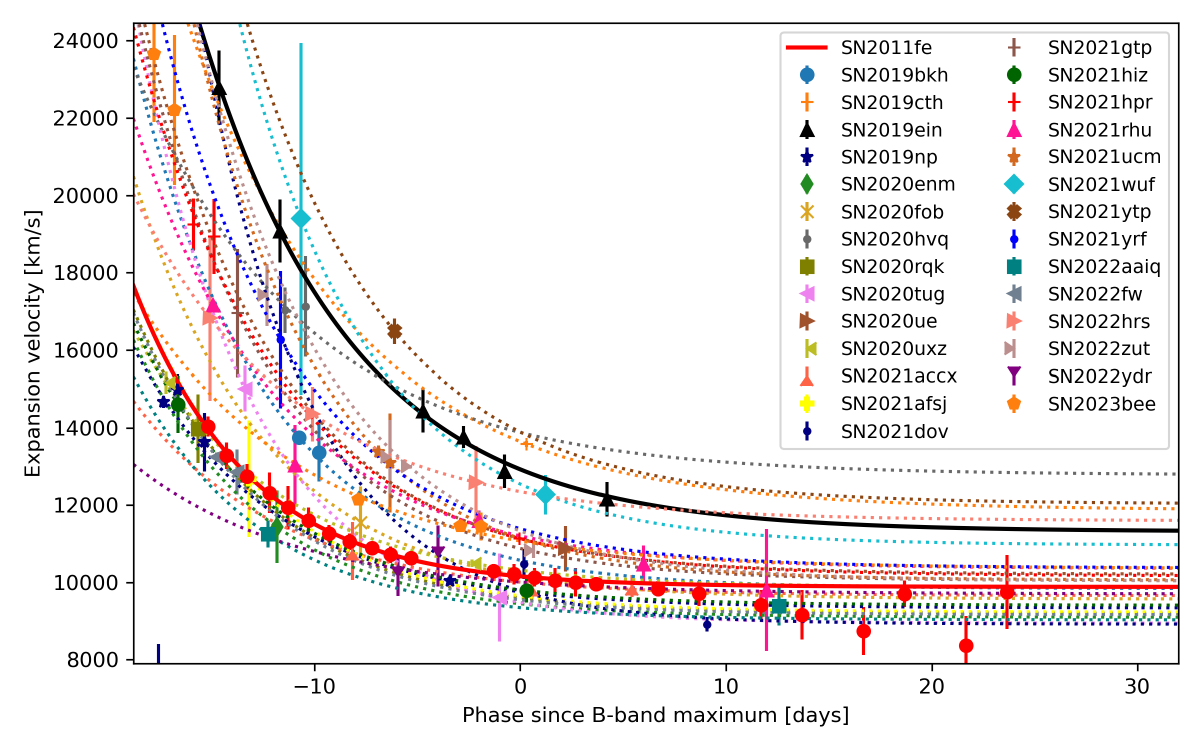}
		\caption{The expansion velocity evolution of the studied objects. Points represent the $v_{\mathrm{exp}}$ values determined from the SiII lines in the spectra, while dotted lines of the same color show the fitted exponential functions. Solid black and red curves show the fitted curves of SN~2019ein and SN~2011fe, respectively.}
		\label{fig:phase_vel}
	\end{figure*}

	Figure~\ref{fig:phase_vel} summarizes the measured velocities (colored symbols) as well as their best-fitting exponential functions (dotted curves). The two solid curves represent SN~2011fe and SN~2019ein. 
	
	\subsection{Classification of the sample SNe into subtypes}
	
	One of the most important parameters of a SN Ia is its color, which is especially interesting in the early phases: it can be a good tracer of the presence of interaction with a companion star \citep{kasen10}, double detonation explosion \citep{polin19}, or $^{56}$Ni-mixing near the surface \citep{magmag20}. 
	
	After correcting for Milky Way and host galaxy extinction, we constructed the
	$(B-V)_0$ color curves of our sample, and estimated the color at the moment of maximum light by fitting a quadratic function to their color evolution.
	
	Many studies have been proposed to further divide SNe Ia into several subclasses based on their spectroscopic and photometric properties; see, e.g., \citet{benetti05, branch06, branch09, wang09, stritzinger18}. The \citet{stritzinger18} classification separates SNe Ia based on their early ($\leq$ 4--5 days after explosion) color evolution: SNe showing $(B-V)_0 \geq 0.2$ mag at around +2 days belong to the early red group, while objects showing $(B-V)_0 < 0.05$ mag are considered as the members of the early blue group. According to \citet{benetti05}, early red SNe~Ia are often associated with Branch Core Normal (CN) and Cool (CL) types showing low velocity gradient (LVG), while early blue objects are usually classified as LVG Shallow Silicon (SS), also known as SNIa-91T subtype. The \citet{wang09} scheme separates SNe~Ia based on their $v_{\rm SiII}$ velocity at maximum light: High Velocity (HV) SNe have $v_{\rm SiII} \gtrsim 12,000$ km\,s$^{-1}$, while Normal Velocity (NV) SNe show $v_{\rm SiII} \approx 10,600$ km\,s$^{-1}$ \citep[see also][for further details and references]{bw17}. The Si II velocity has also been found by \citet{wang13} to be connected to the environment of SNe Ia, with HV events originating from younger, more metal-rich surroundings than NV events. 
	
	By using the estimated $v_{\mathrm{exp}}$ at maximum and the early $(B-V)_0$ color curves, we determined the \citet{wang09} and \citet{stritzinger18} subtypes for the sample SNe where it was possible. Unfortunately, only 7 objects (SNe~2019ein, 2020rqk, 2020tug, 2020uxz, 2021hpr, 2021rhu, 2022fw) had photometry taken at sufficiently early phases to determine their \citet{stritzinger18} group. 5 SNe out of 7 (SNe~2019ein, 2020uxz, 2021hpr, 2021rhu and 2022fw) were found to be consistent with the criteria for the early red subgroup, and the remaining 2 (SNe~2020rqk and 2020tug) belong to the early blue subgroup. Applying the \citet{wang09} classification, most SNe in our sample were found to be of normal velocity (NV) type, and only 6 of them (SNe~2019cth, 2019ein, 2020hvq, 2021wuf, 2021ytp and 2022hrs) belonged to the high velocity (HV) group. It was, however, recently argued that SN~2021hpr may be a transitional object between the Wang NV and HV objects being at the lower end of the HV regime \citep{zhang22}.

	\subsection{Ejecta masses}
	
	Finally, from Equations (\ref{eq:mej}), (\ref{eq:kappa}), and (\ref{eq:ekin}) we inferred the ejecta masses, ejecta opacities and kinetic energies of each SNe Ia in our sample. The results are summarized in Table~\ref{tab:phys_prop}. 
	
	\begin{table*}
		\scriptsize
		\centering
		\caption{Best-fit physical parameters of the sample SNe.}
		\begin{tabular}{lcccccccccc}
			\hline
			SN  &   $t_0$     &   $M_{\mathrm{Ni}}$ & $t_{\mathrm{LC}}$ & $t_{\gamma}$ &   $v_{\mathrm{exp}}$     &   $M_{\mathrm{ej}}$ & $\kappa$ & $E_{\mathrm{kin.}}$ & $M_{\mathrm{Ni}}^{\mathrm{KK19}}$    & Expl.\\ 
			&   (days)     &   ($M_{\odot}$) & (days) & (days)  &   (km\,s$^{-1}$)     &   ($M_{\odot}$) & ($\mathrm{cm}^2$/g) & ($10^{51}$ erg) &   ($M_{\odot}$)  & model  \\ 
			\hline
			SN~2019bkh   &   -16.4 (1.4) &   0.70 (0.05) &   12.40 (1.14) &   43.80 (3.75)   &   10573 (383) &   1.34 (0.33) &   0.09 (0.04) &   0.90 (0.22)  &   0.66 (0.01) &  Both   \\ 
			SN~2019cth   &   -19.4 (0.3) &   0.64 (0.08) &   19.57 (1.21) &   36.64 (1.27)   &   13644 (81) &   1.56 (0.13) &   0.26 (0.05) &   1.75 (0.14)  &   0.46 (0.01) &    Near-M$_{\rm Ch}$  \\ 
			SN~2019ein   &   -15.6 (0.4) &   0.44 (0.05) &   13.41 (0.32) &   32.72 (1.14)   &   12927 (583) &   1.12 (0.18) &   0.16 (0.04) &   1.12 (0.18)  &   0.41 (0.01) &   Both   \\ 
			SN~2019np   &   -17.9 (0.4) &   0.75 (0.05) &   16.21 (1.11) &   41.28 (0.23)   &   9760 (350) &   1.01 (0.08) &   0.20 (0.05) &   0.58 (0.05)  &   0.62 (0.01) &   Sub-M$_{\rm Ch}$   \\ 
			SN~2020enm   &   -13.2 (2.2) &   0.65 (0.06) &   10.93 (1.23) &   40.79 (0.82)   &   9449 (937) &   0.93 (0.22) &   0.09 (0.05) &   0.50 (0.12)  &   0.59 (0.02) &   Sub-M$_{\rm Ch}$   \\ 
			SN~2020fob   &   -12.1 (1.9) &   0.26 (0.05) &   8.26 (1.33) &   37.20 (3.92)   &   10164 (933) &   0.89 (0.35) &   0.06 (0.05) &   0.55 (0.22)  &   0.27 (0.01) &   Both   \\ 
			SN~2020hvq   &   -15.3 (0.5) &   0.40 (0.05) &   10.62 (0.83) &   50.00 (2.33)   &   13881 (945) &   3.01 (0.69) &   0.04 (0.02) &   3.48 (0.80)  &   0.42 (0.01) &   Neither   \\ 
			SN~2020rqk   &   -15.9 (0.1) &   0.59 (0.05) &   12.95 (0.40) &   36.61 (1.17)   &   9602 (887) &   0.77 (0.19) &   0.16 (0.06) &   0.43 (0.11)  &   0.57 (0.02) &   Sub-M$_{\rm Ch}$   \\ 
			SN~2020tug   &   -15.8 (1.0) &   0.70 (0.05) &   11.06 (2.41) &   41.69 (1.75)   &   9485 (864) &   0.98 (0.26) &   0.09 (0.07) &   0.53 (0.14)  &   0.83 (0.01) &   Both   \\ 
			SN~2020ue   &   -12.6 (0.4) &   0.34 (0.05) &   9.57 (0.71) &   34.45 (0.75)   &   11135 (588) &   0.92 (0.14) &   0.09 (0.03) &   0.68 (0.10)  &   0.35 (0.01) &   Sub-M$_{\rm Ch}$   \\ 
			SN~2020uxz   &   -17.7 (0.1) &   0.63 (0.05) &   14.39 (1.00) &   43.60 (2.90)   &   10315 (190) &   1.26 (0.22) &   0.13 (0.04) &   0.81 (0.14)  &   0.59 (0.01) &    Both  \\ 
			SN~2021accx   &   -14.9 (1.0) &   0.66 (0.05) &   11.62 (1.43) &   40.68 (1.97)   &   10008 (438) &   1.04 (0.19) &   0.10 (0.05) &   0.62 (0.12)  &   0.61 (0.01) &   Both   \\ 
			SN~2021afsj   &   -15.6 (0.5) &   0.64 (0.05) &   12.79 (0.95) &   38.28 (1.13)   &   9653 (1502) &   0.85 (0.32) &   0.14 (0.10) &   0.48 (0.18)  &   0.59 (0.01) &   Sub-M$_{\rm Ch}$   \\ 
			SN~2021dov   &   -17.0 (0.6) &   1.19 (0.09) &   14.59 (1.35) &   39.42 (2.03)   &   9779 (286) &   0.93 (0.15) &   0.17 (0.06) &   0.53 (0.09)  &   1.08 (0.01) &  Sub-M$_{\rm Ch}$    \\ 
			SN~2021gtp   &   -15.4 (0.7) &   0.69 (0.06) &   12.80 (1.34) &   37.64 (1.54)   &   10880 (1654) &   1.05 (0.40) &   0.13 (0.10) &   0.74 (0.29)  &   0.66 (0.02) &   Both   \\ 
			SN~2021hiz   &   -16.7 (0.2) &   0.59 (0.05) &   14.98 (0.46) &   37.38 (1.01)   &   9804 (523) &   0.84 (0.14) &   0.20 (0.06) &   0.48 (0.08)  &   0.51 (0.01) &   Sub-M$_{\rm Ch}$   \\ 
			SN~2021hpr   &   -17.0 (0.1) &   0.67 (0.05) &   13.97 (0.27) &   39.96 (0.68)   &   11132 (560) &   1.24 (0.17) &   0.14 (0.03) &   0.92 (0.12)  &   0.62 (0.01) &   Both   \\ 
			SN~2021rhu   &   -14.5 (0.1) &   0.45 (0.05) &   13.81 (0.73) &   30.64 (2.79)   &   11206 (477) &   0.74 (0.20) &   0.22 (0.09) &   0.56 (0.15)  &   0.37 (0.01) &   Sub-M$_{\rm Ch}$   \\ 
			SN~2021ucm   &   -16.1 (3.2) &   0.76 (0.12) &   13.94 (3.50) &   42.58 (2.98)   &   11137 (719) &   1.41 (0.38) &   0.12 (0.10) &   1.05 (0.28)  &   0.70 (0.07) &   Both   \\ 
			SN~2021wuf   &   -15.1 (0.9) &   0.72 (0.09) &   12.33 (1.59) &   35.09 (1.47)   &   12547 (2528) &   1.21 (0.59) &   0.12 (0.12) &   1.14 (0.56)  &   0.56 (0.01) &   Both   \\ 
			SN~2021yrf   &   -15.0 (0.6) &   0.55 (0.05) &   11.91 (0.78) &   37.27 (1.53)   &   11391 (1771) &   1.13 (0.44) &   0.11 (0.08) &   0.88 (0.34)  &   0.52 (0.01) &   Both   \\ 
			SN~2021ytp   &   -13.1 (0.8) &   0.64 (0.06) &   11.28 (1.08) &   27.33 (1.69)   &   13894 (328) &   0.90 (0.15) &   0.15 (0.06) &   1.04 (0.18)  &   0.56 (0.05) &   Sub-M$_{\rm Ch}$   \\ 
			SN~2022aaiq   &   -14.5 (1.2) &   0.49 (0.06) &   9.02 (1.59) &   53.29 (3.06)   &   9347 (349) &   1.55 (0.29) &   0.04 (0.02) &   0.81 (0.15)  &   0.52 (0.01) &   Near-M$_{\rm Ch}$   \\ 
			SN~2022fw   &   -16.4 (0.6) &   0.71 (0.07) &   12.59 (1.26) &   45.54 (2.11)   &   9551 (355) &   1.18 (0.20) &   0.10 (0.04) &   0.65 (0.11)  &   0.69 (0.02) &   Both   \\ 
			SN~2022hrs   &   -15.3 (0.2) &   0.60 (0.05) &   12.37 (0.35) &   39.29 (1.31)   &   12349 (1257) &   1.47 (0.40) &   0.10 (0.04) &   1.35 (0.36)  &   0.58 (0.01) &   Both   \\ 
			SN~2022ydr   &   -13.9 (1.0) &   0.24 (0.05) &   13.02 (1.28) &   27.95 (1.60)   &   10010 (649) &   0.49 (0.12) &   0.27 (0.14) &   0.29 (0.07)  &   0.19 (0.01) &   Neither   \\ 
			SN~2022zut   &   -20.3 (1.6) &   0.61 (0.06) &   15.14 (1.72) &   47.88 (4.55)   &   11332 (497) &   1.84 (0.51) &   0.11 (0.06) &   1.42 (0.39)  &   0.62 (0.03) &   Near-M$_{\rm Ch}$   \\ 
			SN~2023bee   &   -17.1 (0.5) &   0.82 (0.05) &   13.25 (0.79) &   45.14 (1.61)   &   11074 (829) &   1.56 (0.35) &   0.10 (0.04) &   1.15 (0.25)  &   0.79 (0.02) &   Near-M$_{\rm Ch}$   \\ 
			
			\hline
		\end{tabular}
		\tablenotetext{}{The first 4 columns list the best-fit parameters given by Minim, then the expansion velocities, the ejecta masses, the calculated opacities, kinetic energies (see Section~\ref{sec:phys_par}), nickel masses from the method of \citet{kk19}, and the consistency with near-Chandra and/or sub-Chandra explosion models are shown.}
		\label{tab:phys_prop}
	\end{table*}

	\section{Discussion}    \label{sec:discussion}
	
	In this section we present constraints and correlations between the physical parameters and progenitor systems of the studied SNe Ia, and discuss their implications for the explosion mechanisms.
	
	\subsection{Comments on individual objects}
	\label{sec:individual}
	
	In this subsection we discuss and compare our results with those obtained by others for some individual SNe in our sample. The objects listed here are the ones that either had published results that were comparable to our work, or they were found to have some peculiarities during our analysis.
	
	\begin{itemize}
		\item \underline{SN~2019ein}: \\
		We classified SN~2019ein as a high velocity SN Ia, consistent with \citet{kawabata20} and \citet{xi22}. Additionally, \citet{xi22} estimated a nickel mass of $0.27-0.31$ M$_{\odot}$, which is lower than our result of $0.44 \pm 0.05$ M$_{\odot}$. The reason for the disagreement is most likely the $\sim 0.4$ mag lower distance modulus adopted in the previous papers. \citet{xi22} also proposed that SN~2019ein might be the result of a sub-Chandra double detonation explosion. \citet{pellegrino20} also estimated a nickel mass of $0.33$ M$_{\odot}$, which, again, is lower than our result, but it was based on spectral modeling instead of photometry. Contrary to \citet{xi22}, they argue that this event is a product of the delayed-detonation scenario. Our estimated ejecta mass for SN~2019ein is $1.12 \pm 0.18$ M$_{\odot}$, which is consistent with a sub-Chandra explosion, but because of the relatively high uncertainty of the ejecta mass, the possibility for a Chandrasekhar-mass delayed-detonation explosion cannot be excluded.
		
		\item \underline{SN~2019cth}: \\
		SN~2019cth was a 91T-like event, but with low maximum luminosity and low nickel mass, contrary to the expectations. The reason for the inconsistency could be due to the poor phase coverage of our data around maximum light. The uncertainty of the nickel mass is also significant: the value given by the Minim fitting ($0.64 \pm 0.08$ M$_{\odot}$) is significantly higher than the one calculated by the method of \citet{kk19} ($0.46 \pm 0.01$ M$_{\odot}$). Thus, the parameters for this object listed in Table~\ref{tab:phys_prop} should be considered with caution.
		
		\item \underline{SN~2019np}: \\
		\citet{sai22} also applied the Minim code and the Arnett-model to fit the quasi-bolometric light curve of SN~2019np, which resulted in a nickel mass of $0.66 \pm 0.05$ M$_{\odot}$. This value is slightly lower than our $M_{\rm Ni} = 0.75 \pm 0.05$ M$_{\odot}$. From spectroscopy \citet{sai22} estimated the ejecta velocity as $\sim 10,200$ km~s$^{-1}$ at B-band maximum, which agrees with our value within the errors. On the other hand, they estimated a high ejecta mass of $\sim 1.34 \pm 0.12$ M$_{\odot}$ compared to our value of $\sim 1.01 \pm 0.08$ M$_{\odot}$. However, inserting their reported $v_{\mathrm{exp}}$ and $t_{\gamma}$ values into Equation (\ref{eq:mej}), we get an ejecta mass of $\sim 0.9$~M$_{\odot}$. Thus, we believe that the ejecta mass reported by \citet{sai22} is an overestimate, if we use their parameters at face value. They also found extra light in the early phases of the light curve, and attribute it to nickel-mixing. Our data were not taken early enough to confirm their finding. 
		
		\item \underline{SN~2020hvq}: \\
		This SN appeared right in the dust lane of the edge-on galaxy UGC~10561 (see Figure~\ref{fig:stamps1}).
		Our data on SN~2020hvq extend well beyond +80 days, but the Minim code could not fit the tail beyond +60 days well. This may have resulted in an overestimate of the $t_{\gamma}$ value. The best-fit $t_{\gamma} \sim 50$ days is not extremely long, but coupled with the relatively high expansion velocity, $v_{\rm exp} \approx 13,900$ km~s$^{-1}$, gives a very high ejecta mass of $M_{\rm ej} \approx 3$ M$_\odot$. Such a high ejecta mass is inconsistent with not only the current explosion models for SNe Ia, but also with the $t_{LC}$ parameter ($\sim 10.6$ d) that would imply $M_{\rm ej} \approx 1.2$~M$_\odot$ according to the method by KTR20.
		The long $t_\gamma$ could be due to some extra light in the late-phase photometry, either by the insufficient subtraction of host galaxy background from our frames, a light echo on significant dust content in the host around the line of sight, or a late-time circumstellar interaction \citep{terwel24}. Because SN~2020hvq is a significant outlier in our sample, it was excluded from further analysis.

		\item \underline{SN~2020tug}: \\
		Similar to SN~2019cth, this object lacked near-maximum observations in our data, thus, the estimated nickel mass may be more uncertain.
		
		\item \underline{SN~2021dov}: \\
		SN~2021dov is a 91T-like SN~Ia, which showed the highest maximum luminosity in our sample ($\sim 2\times 10^{43}$~erg\,s$^{-1}$). As expected, this implies a high nickel mass of $\sim 1.2$~M$_{\odot}$. However, the inferred ejecta mass was found to be unexpectedly low, $0.93 \pm 0.15$~M$_{\odot}$. Since the condition $M_{\rm Ni} > M_{\rm ej}$ is unphysical, we suspect that in this case some of the initial assumptions, probably the centrally located powering source, of the Arnett-model fails (see Section~\ref{sec:caveats} for further discussion). Significant interaction with a H-poor CSM that increases the maximum luminosity beyond the radioactive heating, or a non-spherically expanding ejecta are also possible explanations. Note that KTR20 also found a similar outlier (SN~2017erp) in their sample, which showed $M_{\rm Ni} \approx M_{\rm ej}$. 
		
		\item \underline{SN~2021hpr}: \\
		\citet{lim23} observed excess light in the early LC, and attributed this to the shock-heated cooling emission (SHCE) that arises from interaction with a companion star. Using the SHCE model along with delayed-detonation models, they were able to fit the early LC adequately. While the double-detonation model of \citet{polin19} also predicts extra light in the early LC, \citet{lim23} found that it does not fit the observations as well as the SHCE model. SN~2021hpr is the third SN Ia discovered in NGC 3147 during the last half-century. A study of SN~2021hpr and its siblings were presented by \citet{barna23}, who also used the data from the RC80 and BRC80 telescopes, but performed their own independent analysis. Their estimated nickel mass ($0.44 \pm 0.14$~M$_{\odot}$) is lower than the value given in Table~\ref{tab:phys_prop} ($0.67 \pm 0.05$~M$_{\odot}$), but their expansion velocity ($11200 \pm 1200$ km\,s$^{-1}$), and ejecta mass ($1.12 \pm 0.28$~M$_{\odot}$) are both in agreement with our values (see Table~\ref{tab:phys_prop}). The disagreement in the nickel mass can be traced back to their lower adopted extinction and distance modulus. 
		
		\item \underline{SN~2021rhu}: \\
		SN~2021rhu was a sub-luminous and very red event near maximum compared to the other SNe in our sample. According to \citet{harvey23}, it may represent a transitional object between normal and 91bg-like SNe Ia, i.e., a 86G-like object. By modeling its spectra, \citet{harvey23} found that a delayed-detonation scenario is more likely than a double-detonation explosion. In this paper, we estimate a nickel mass of $\sim 0.45 $~M$_{\odot}$, which is close to the calculated $\sim 0.74 $~M$_{\odot}$ ejecta mass. The low ejecta mass is a consequence of the low $t_{\gamma}$ value ($\sim31$ days), which is one of the lowest in our sample. Since the low $t_{\gamma}$ and $M_{\mathrm{ej}}$ values seem to be relatively well-constrained and consistent with SN~2021rhu being a lower mass object, we speculate that the too low ejecta mass might be due to the failure of the Arnett-model, e.g., because of the unusual Ni-distribution (see Section~\ref{sec:caveats}).

		\item \underline{SN~2022hrs}: \\
		A SALT3 fitting of the $BVri$ LC of SN~2022hrs taken by 0.4m LCO telescopes was recently published by \citet{risin23}. They got significantly lower Ni-mass ($\sim 0.16$ M$_\odot$) compared to our result ($\sim 0.6$ M$_\odot$, Table~\ref{tab:phys_prop}). The disagreement is probably due to their different methodology: they used the best-fit LC amplitude provided by SALT3 to estimate the Ni-mass, while our result was inferred from the quasi-bolometric luminosities. Nevertheless, their result seems to be an underestimate compared the usual range of nickel masses in SNe Ia (0.3 -- 1.1 M$_\odot$).

		\item \underline{SN~2022ydr}: \\
		Our observations of SN~2022ydr lack adequate coverage in the pre-maximum period, and the observations only extend until $\sim 30$ days after maximum. This can cause an uncertain fitting of the $t_{\gamma}$ parameter, which in turn may lead to the low ejecta mass estimate of $\sim 0.49 \pm 0.12$ M$_\odot$, making it the only object in our sample that lies significantly below the sub-Chandra explosion models. SN~2022ydr was of a  Ia-91bg subtype, and accordingly it had the lowest maximum luminosity ($\sim 0.55\times 10^{43}$ erg\,s$^{-1}$), which is reflected in the relatively low ($\sim 0.24$ M$_\odot$) nickel mass with respect to the rest of the sample. Compared to most 91bg-like SNe, the Ni-mass of SN~2022ydr is a bit higher, but not unprecedented. For example, \citet{sullivan11} used Arnett's rule to estimate $\sim 0.2$ M$_\odot$ of $^{56}$Ni for SN~2007au that had a peak luminosity similar to that of SN~2022ydr. It is possible that these higher-mass 91bg-like SNe Ia belong to a transitional group between 91bg and normal SNe Ia \citep{li22}. Another possibility is that at such low nickel masses the Arnett model might overestimate M$_{\rm Ni}$.

		\item \underline{SN~2023bee}: \\
		SN~2023bee was found to have excess blue light in the early part of the light-curve \citep{hosseinzadeh23}, which is most probably due to companion interaction \citep{kasen10}.
		
	\end{itemize}

	\subsection{Comparison with other works} \label{sec:comparison}
	In this subsection, we compare our methods and results to those given by previous publications.
	
	\subsubsection{Comparison with \texorpdfstring{\citet{kk19}}{Khatami \& Kasen (2019)}}
	
\begin{figure}[H]
		\centering
		\includegraphics[width=1.0\linewidth]{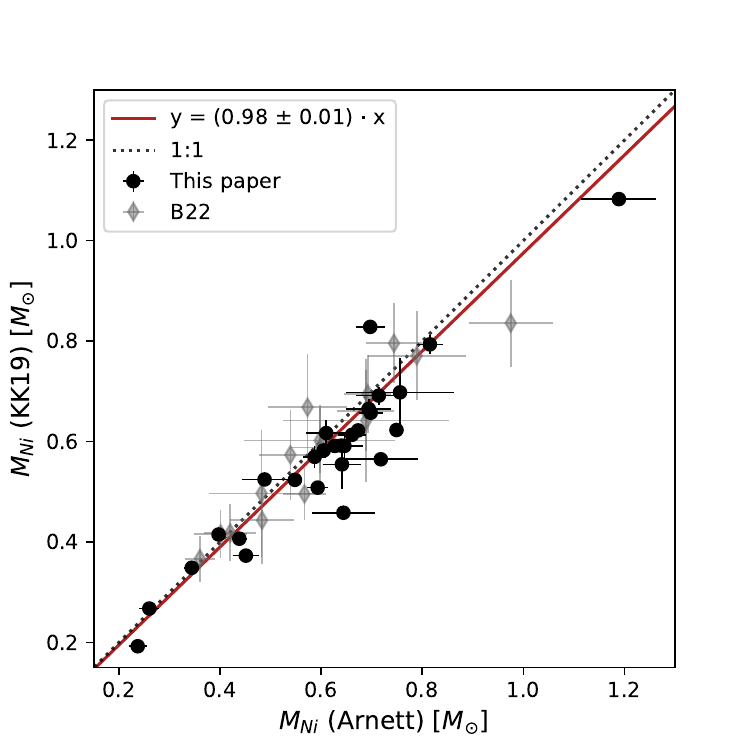}
		\caption{The comparison of the nickel masses estimated by fitting the bolometric light-curves with the Arnett-model 
			(x-axis) to those calculated with the method of \citet{kk19} 
			(y-axis). Grey diamonds show the results by \citet{bora22}, while SNe from the current sample are plotted with black circles. The solid red line indicates the linear relation fitted to all data, which is close to the 1:1 relation (shown by black dotted line).}
		\label{fig:kk19_comp}
	\end{figure}

	Instead of fitting the entire bolometric LC, \citet{kk19} used only the maximum bolometric luminosity and the LC rise time to constrain the nickel mass. By using their method, we re-calculated the $^{56}\mathrm{Ni}$ masses for the objects in our sample.
	Comparison of the nickel masses derived with the two different methods can be seen in Figure~\ref{fig:kk19_comp}, which shows that the two methods give similar results, consistently with \citet{bora22}. The consistency between the results from two different approaches that constrain the initial Ni-mass suggests that systematic errors of the estimated Ni-masses are not severe. It may strengthen the validity and applicability of the Arnett-model for most of the SNe Ia in the present sample.
	
	\begin{figure*}
		\centering
		\includegraphics[width = 0.9\textwidth]{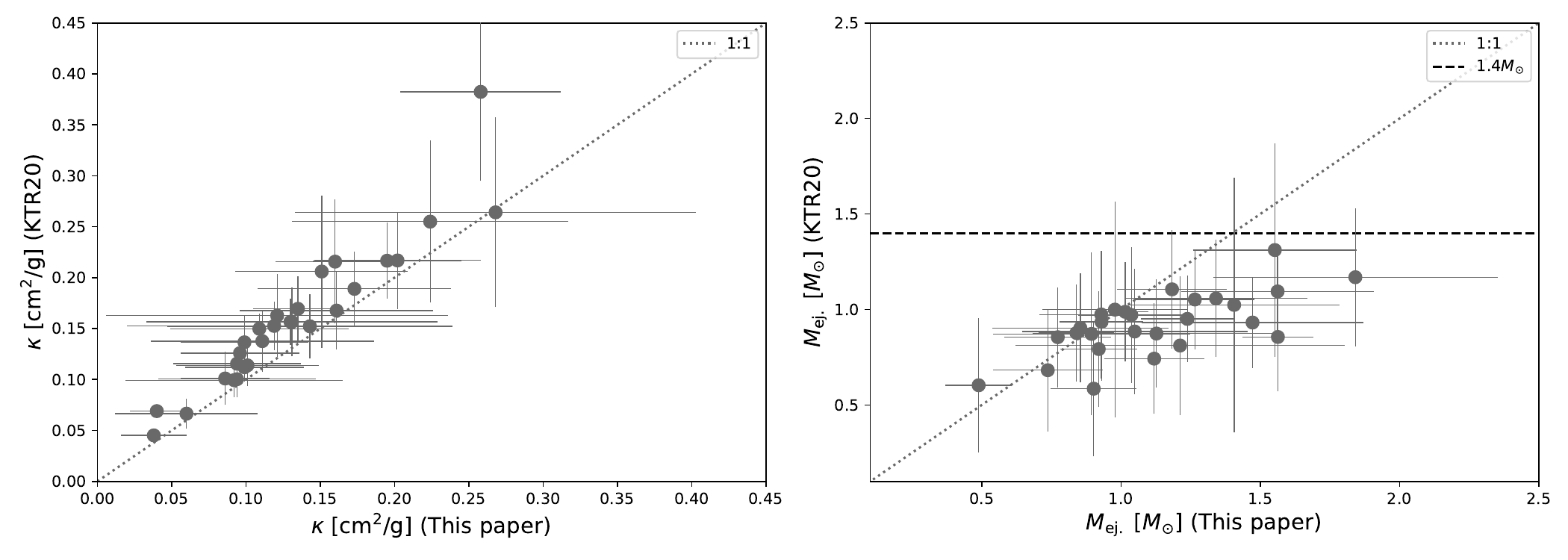}
		\caption{Comparison of ejecta opacities and masses with those obtained with the method of KTR20. 1:1 relations are shown with dotted lines, while the dashed line indicates the Chandrasekhar mass. Note that the method of KTR20 uses an assumption of $M_{\rm ej} \leq M_{Ch}$, and this could be the cause for the deviation from the 1:1 relation in the $M_{\rm ej} > 1$~M$_\odot$ mass range.}
		\label{fig:ktr_bzs_comp}
	\end{figure*}

	\begin{figure*}
		\centering
		\includegraphics[width = 0.9\textwidth]{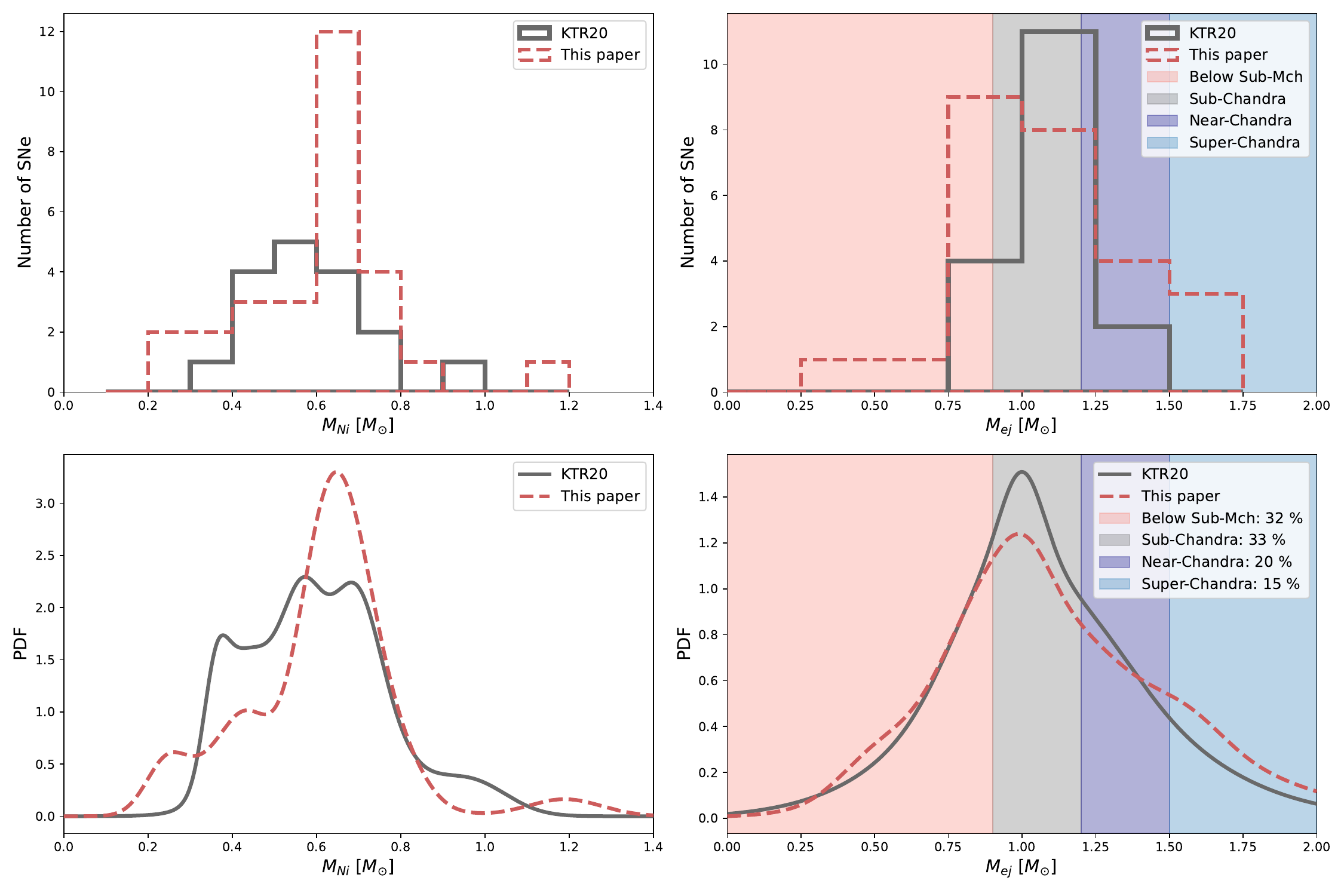}
		\caption{Top panels: comparison of the histograms of the nickel and ejecta mass distributions of our sample (red) with that of KTR20 (grey). 
			Bottom panels: The sums of individual Gaussian distributions of the nickel and ejecta masses with their errors taken into account as widths of the individual Gaussians. The colored background in the right panels codes the different mass regimes for super-Chandra (M$_{\mathrm{ej}}$ $> 1.5$ M$_\odot$), near-Chandra ($1.2$ M$_\odot <$ M$_{\mathrm{ej}}$ $< 1.5$ M$_\odot$), sub-Chandra ($0.9$ M$_\odot <$ M$_{\mathrm{ej}}$ $< 1.2$ M$_\odot$) and below sub-Chandra (M$_{\mathrm{ej}}$ $< 0.9$ M$_\odot$) ejecta masses.   
		}
		\label{fig:ktr_bzs_comp_hist}
	\end{figure*}
	
	\begin{figure*}[ht]
		\centering
		\includegraphics[width = 0.9\textwidth]{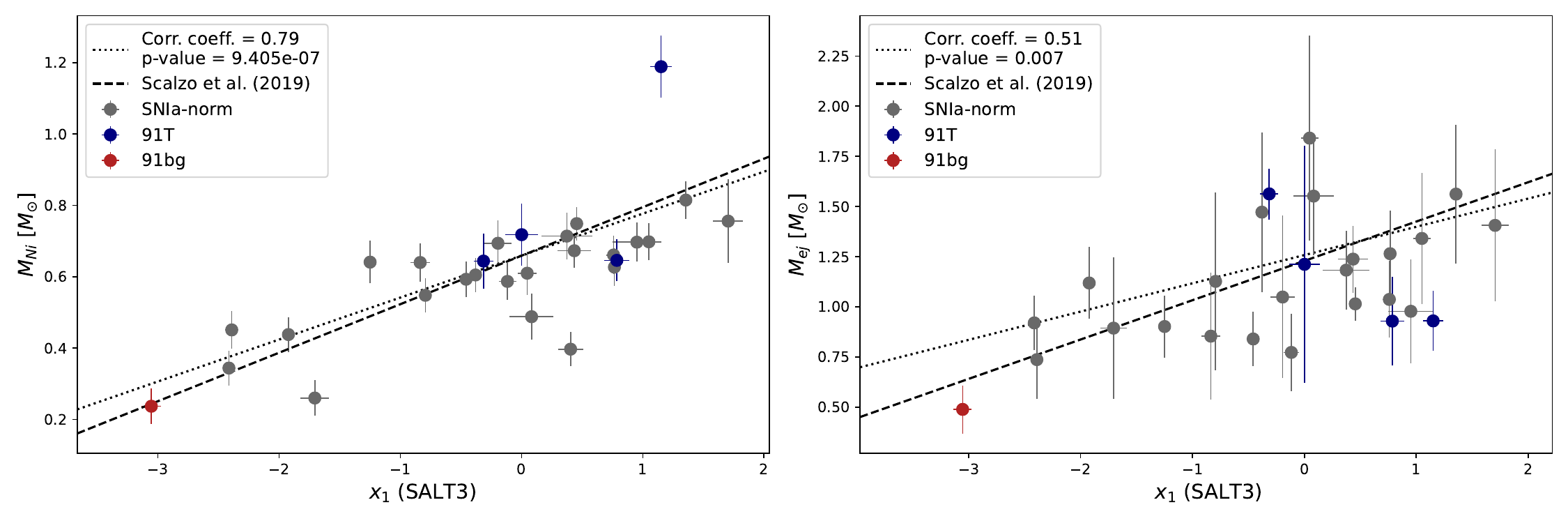}
		\caption{Nickel mass (left) and ejecta mass (right) plotted against the SALT3 $x_1$ parameter. The relations determined by \citet{scalzo19} are shown with dashed lines, while dotted lines present the relations fitted by us. 91T-like SNe are marked with blue dots, while grey and red symbols denote the normal and 91bg-like SNe, respectively.}
		\label{fig:salt3_x1}
	\end{figure*}

	\subsubsection{Comparison with KTR20}
	
	As mentioned in Section~\ref{sec:phys_par}, in the present paper we estimated the physical parameters of the studied SNe Ia similar to KTR20, except that the expansion velocities were inferred from the Doppler shift of the Si~II $\lambda$6355 feature in the available spectra. For comparison, we also derived the physical parameters for the whole sample with the method of KTR20, i.e., without using spectroscopic velocities. 
	After comparing the two sets of parameters taken with and without spectroscopic velocities, 
	we found that the $v_{\rm exp}$ and $E_{\rm kin}$ parameters are significantly different. This is not surprising given the very approximate methodology used by KTR20 for estimating the velocities without spectroscopy. Thus, our new velocities and kinetic energies are more reliable than those obtained by KTR20 for their sample. Note also that \citet{scalzo19} revealed practically no correlation between their expansion velocities (inferred directly from the kinetic energy provided by explosion models) and $v_{\rm SiII}$, thus, they concluded that $v_{\rm SiII}$ is not a good proxy for estimating the kinetic energy of the bulk ejecta. 
	
	On the other hand, the ejecta masses ($M_{\rm ej}$), and opacities ($\kappa$) given by the two methods are in good agreement, as seen in Figure~\ref{fig:ktr_bzs_comp}. We also compare the nickel masses and ejecta masses between our sample and that of KTR20 in Figure~\ref{fig:ktr_bzs_comp_hist} (note that in Figure~\ref{fig:ktr_bzs_comp} the objects are the same, but the methods are different, while in Figure~\ref{fig:ktr_bzs_comp_hist} we show the results of two different methods being applied to two different samples).
	
	Figure~\ref{fig:ktr_bzs_comp_hist} demonstrates that the two samples have similar distributions of both the ejecta masses and the nickel masses, even though different methods were applied to different samples. This shows that in spite of not using spectral information, the ejecta masses estimated by KTR20 are realistic, and their method is useful for estimating the ejecta mass in cases where getting near-maximum spectra is not possible. It is important to note that in their analysis, KTR20 set an upper limit of $1.4~$M$_{\odot}$ for the ejecta mass and a lower limit of 10000 km\,s$^{-1}$ for $v_{\mathrm{exp}}$. Our method does not have these limits, so the distribution of ejecta masses can be wider, as seen in Figure~\ref{fig:ktr_bzs_comp_hist}. Other than that, we found good agreement between the two distributions: SNe in the KTR20 sample were found to have a mean $M_{\rm ej}$ of $\sim 1.07$~M$_\odot$ and standard deviation of $0.15$~M$_{\odot}$, which is consistent with those of our sample, $\sim 1.12$~M$_{\odot}$ and $0.30$~M$_{\odot}$, respectively.

	\subsubsection{Comparison with \texorpdfstring{\citet{scalzo19}}{Scalzo et al. (2019)}}
	\label{sec:scalzo19}
	
	Previously, ejecta masses for SNe~Ia were also inferred by \citet{scalzo10}, \citet{scalzo12}, and \citet{scalzo14a}. They used a method that was purely photometric, without any spectroscopic information, similar to KTR20. Their theoretical approach was based on a combination of several theoretical Ia explosion models, and they applied a different formalism than KTR20 and we did in the present paper. \citet{scalzo14a} estimated the $^{56}$Ni and ejecta masses for their sample containing 19 SNe~Ia, and concluded that about 50\% of their sample SNe arose from the explosion of sub-Chandra WDs \citep[see also][]{scalzo14b}. Their method was further applied on a larger sample ($\sim45$ SNe) and reached a similar conclusion \citep{scalzo19}.

	Taking into account the uncertainties, 
	22 out of 27 SNe (81 \%) in our sample are consistent with sub-Chandra WD masses, but 13 of them are also consistent with both sub-Chandra and near-Chandra masses (see Table~\ref{tab:phys_prop}). This leaves 9 SNe (33 \%) in our sample whose ejecta masses are consistent with only sub-Chandra WDs. If we take the inferred ejecta masses at their face value, 11 out of 27 SNe (41 \%) fall within the mass range of sub-Chandra explosion, i.e., between 0.9 and 1.2~M$_\odot$. This is
	somewhat lower than the value found by \citet{scalzo19}, but the uncertainties of the inferred ejecta masses, or the minor differences between the methodologies could explain the differences.
	Nevertheless, our results are consistent with the previous conclusions that a significant fraction ($\sim 50$\%) of SNe Ia produce sub-Chandra ejecta masses after explosion. 
	
	\begin{figure*}[ht]
		\centering
		\includegraphics[width = 0.9\textwidth]{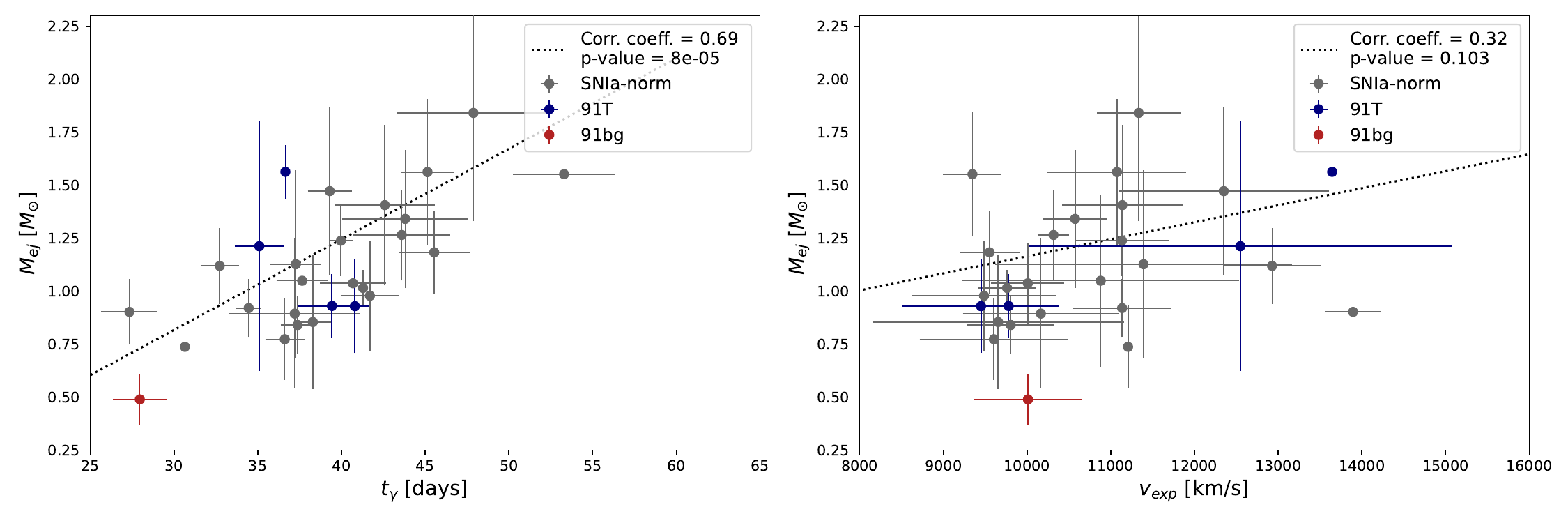}
		\caption{Ejecta masses plotted against gamma-ray timescales (left) and expansion velocities (right) for all observed SNe. The dotted lines show the relations fitted by us, blue dots mark the 91T subtypes, grey dots the normal SNe~Ia, and red dots show the 91bg subtypes.}
		\label{fig:tg_mej}
	\end{figure*}

	\subsection{Correlations between the parameters} \label{sec:corr}
	
	In this section we examine the correlation between the inferred physical parameters using p-values and Pearson correlation coefficients, which are presented in Figures \ref{fig:salt3_x1}, \ref{fig:tg_mej} and \ref{fig:color_lmax_models}.
	In Figure~\ref{fig:salt3_x1} we show the correlation between the SALT3 $x_1$ parameter and the initial nickel mass (left panel) and the ejecta mass (right panel). The best-fitting linear equations are 
	\begin{equation}
		M_{\mathrm{Ni}} = 0.117 ( \pm 0.022) \cdot x_1 + 0.659 (\pm 0.027),
	\end{equation}
	and
	\begin{equation}
		M_{\mathrm{ej}} = 0.140 (\pm 0.057) \cdot x_1 + 1.258 (\pm 0.055).
	\end{equation}
	
	Both of these fittings agree well with those of \citet{scalzo19} within the errors, and show that the light curve decline rate (measured by the SALT3 $x_1$ parameter) correlates with the initial mass of the radioactive nickel synthesized in the explosion, as well as with the ejecta mass: lower decline rates (more positive $x_1$) correspond to higher $M_{\rm Ni}$ and $M_{\rm ej}$.
	
\begin{figure}[ht]
		\centering
		\includegraphics[width = 0.9\linewidth]{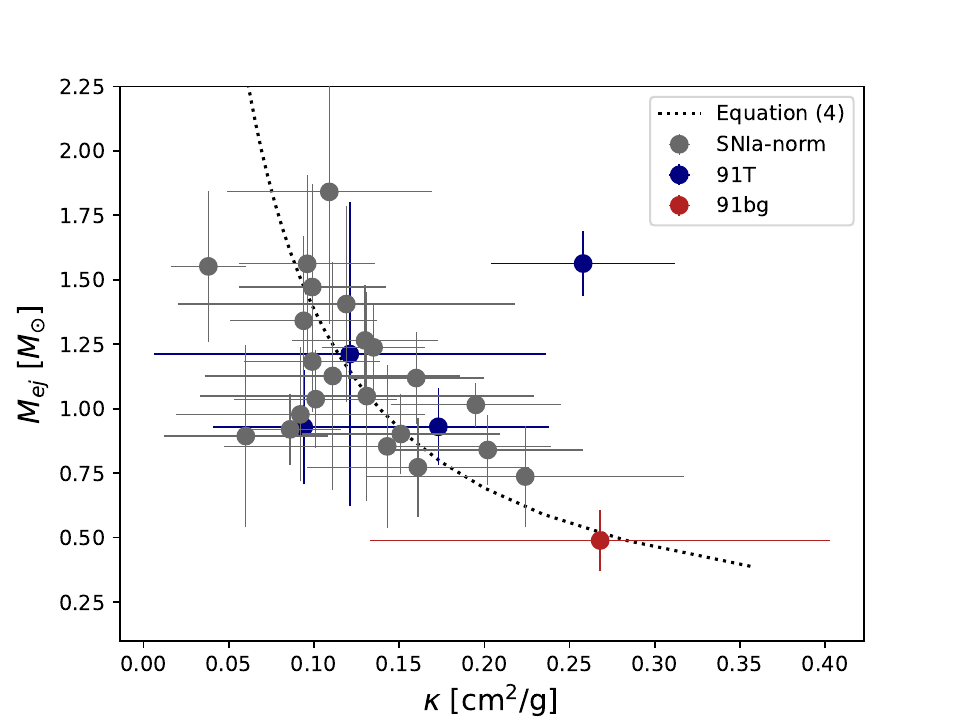}
		\caption{Ejecta masses plotted against the optical opacity $\kappa$. The dotted line shows the prediction by Equation~(\ref{eq:kappa}), blue dots mark the 91T subtypes, grey dots the normal SNe-Ia, and red dots show the 91bg subtypes.}
		\label{fig:kappa_mej}
	\end{figure}

	Similar correlation can be seen between $M_{\mathrm{ej}}$ and the timescale of the gamma-ray leakage ($t_{\gamma}$), as shown in Figure~\ref{fig:tg_mej} (left panel). This relation illustrates that in more massive ejecta it takes longer for the gamma-photons to diffuse out, therefore $t_\gamma$ is a good indicator of progenitor mass (see also KTR20, but see Section~\ref{sec:caveats} for caveats). The correlation with the expansion velocities (Figure~\ref{fig:tg_mej} right panel) is weaker. From Equation~(\ref{eq:mej}), $M_{\rm ej} \approx v_{\rm exp}^2$ is expected, but, because of the presence of the $M_{\rm ej} \approx t_\gamma^2$ dependency, the correlation with the velocities looks smeared. In Figure~\ref{fig:kappa_mej} we show the relation between the ejecta mass and the optical opacity. The dotted line indicates the expected dependency from Equation~(\ref{eq:kappa}) using the mean values for $t_{\rm LC}$ and $v_{\rm exp}$ (12.81 days and 10,929 km~s$^{-1}$, respectively) inferred from our sample.
	
	\subsection{Comparison with explosion models}
	\label{sec:explosion}

	In this subsection we examine if our results are consistent with recently published explosion models. 
	
\begin{figure}
		\centering
		\includegraphics[width=1.1\linewidth]{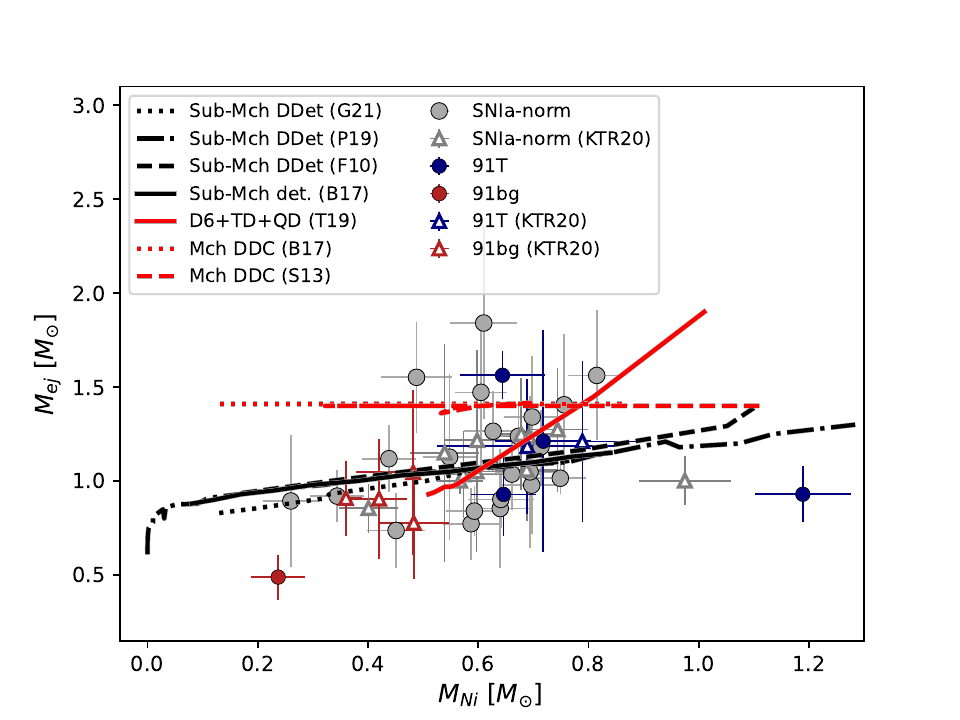}
		\caption{Comparison of the distribution of the observed SNe on the $M_{\mathrm{Ni}}$-$M_{\mathrm{ej}}$ diagram with the predictions of various explosion models \citep{fink10, seitan13, blondin17, tanikawa19, polin19, gronow21}. The blue, grey and red colors code the 91T-, normal- and 91bg subtypes respectively. SNe from this paper are plotted with filled circles, while hollow circles represent the objects from KTR20.}
		\label{fig:mni_mej_models}
	\end{figure}

	First, we compare our estimated ejecta and nickel masses to the predictions of various detonation models in Figure~\ref{fig:mni_mej_models}. Here we show the relations predicted by He-shell double detonation models: the simulations by \citet{gronow21} (dotted line), the parameter survey of \citet{polin19} (dash-dotted line), the hydrodynamic simulations of \citet{fink10} (dashed line), and the radiative transfer simulations of pure central detonations in sub-$M_{\mathrm{Ch}}$ carbon–oxygen WDs by \citet{blondin17} (solid line). We also compare our results with the predictions of the $\mathrm{D}^6$, TD (triple detonation), and QD (quadruple detonation) models of \citet{tanikawa19} (red solid line), and the delayed-detonation models of $M_{\rm Ch}$ WDs by \citet{blondin17} (red dotted line), and \citet{seitan13} (red dashed line).
	
	As Figure~\ref{fig:mni_mej_models} illustrates, the majority of objects in our observational sample (22 out of 27) are consistent with the current sub-$M_{\mathrm{Ch}}$ models, including the thin He-shell double detonation models in single-degenerate progenitors. Even though some of the calculated ejecta masses are below 0.9~M$_\odot$, which is the low-mass cutoff of the current sub-Chandra models that produce enough $^{56}$Ni to be consistent with observed SNe Ia, the uncertainties of the inferred ejecta masses still make them consistent with the models (i.e., their difference from the cutoff mass is less than $1 \sigma$). 
	Moreover, the DD models of \citet{tanikawa19}, involving two merging WDs in a double-degenerate configuration, are also consistent with the observations, at least in the nickel mass range of 0.5-0.7 $M_{\odot}$, but this is a bit uncertain since those models do not extend significantly below $\sim 1$ $M_\odot$ ejecta mass, unlike the SNe in our observational sample. The \citet{tanikawa19} models above $\sim 1~M_{\odot}$ are TD and QD explosions, where the companion WD also explodes, resulting in higher ejected masses. Overall, sub-Chandra detonations as well as double degenerate models seem to be consistent with the majority of our observational sample.
	
	There is only 1 object (the 91bg-like SN~2022ydr) that seems to be significantly below the cutoff mass, i.e., inconsistent with the current explosion models. Since all of these events are spectroscopically confirmed SNe Ia, this inconsistency is likely due to either some difficulties with the observations, or the failure of the Arnett-model for this particular object (see Section~\ref{sec:caveats}).
	
	On the other hand, Figure~\ref{fig:mni_mej_models} shows that more than 50\% of the SNe (17 out of 27) also fall within the mass range of near-Chandra DDC models ($1.2 \lesssim M_{\rm ej} \lesssim 1.5$~M$_\odot$) if their errorbars are taken into account. However, only 4 of them exceed 1.2 M$_\odot$ significantly, as the others have uncertainties that extend below 1.2 M$_\odot$. The near-Chandra models, plotted with red dashed and dotted lines in Figure~\ref{fig:mni_mej_models}, have the same $M_{\rm ej} \approx 1.44$~M$_\odot$ in general, and they show practically no correlation between their $M_{\rm ej}$ and $M_{\rm Ni}$. 
	
	It is important to note that due to the uncertainties of the calculated ejecta masses, a significant fraction of our sample (13/27, 48\%) are consistent with both the near-Chandra and sub-Chandra explosion scenarios, so it is not possible to distinguish between these two possibilities in those cases.

	These fractions can also be estimated using the Gaussian Probability Density Function (PDF) shown in the bottom panel of Figure~\ref{fig:ktr_bzs_comp_hist}. Integrating the PDF below 0.9~M$_\odot$, between 0.9 -- 1.2~M$_\odot$, between 1.2 -- 1.5~M$_\odot$ and above 1.5~M$_\odot$, corresponding to below-sub-Chandra, sub-Chadra, near-Chandra and super-Chandra mass regimes, respectively, we got the probabilities for these regimes as 0.32, 0.33, 0.20 and 0.15, respectively. These are also indicated in the legend of Figure~\ref{fig:ktr_bzs_comp_hist}. Again, since the uncertainties seem to be higher than the mass difference between the sub-Chandra, and near-Chandra channels, it is difficult to draw a definite conclusion, but it seems to be slightly more probable (at least in the present sample) that a SN belongs to the sub-Chandra mass regime instead of being a near-Chandra event.

	As seen in Figure~\ref{fig:mni_mej_models}, the range of the inferred Ni-masses is fully consistent with the ones predicted by both the near-Chandra and sub-Chandra models. For the former, where the ejecta masses are the same (M$_{\rm Ch}$), the differences in the Ni-masses can be achieved by the number of ignition sparks and the implementation of the deflagration-to-detonation transition \citep[e.g.,][]{seitan13}. On the contrary, in sub-Chandra explosions the synthesized Ni-mass is connected with the mass of the WD (and partly with the details of the ignition mechanism), which is reflected by the slope of the lines representing the sub-Chandra models in Figure~\ref{fig:mni_mej_models}. The fact that most of the inferred ejecta masses also seem to show a similar trend suggests that those SNe (or at least some of them) might also be produced by sub-Chandra explosions, but the uncertainties of the inferred ejecta masses prevent making a definite conclusion at the moment.

\begin{figure}
		\centering
		\includegraphics[width=1\linewidth]{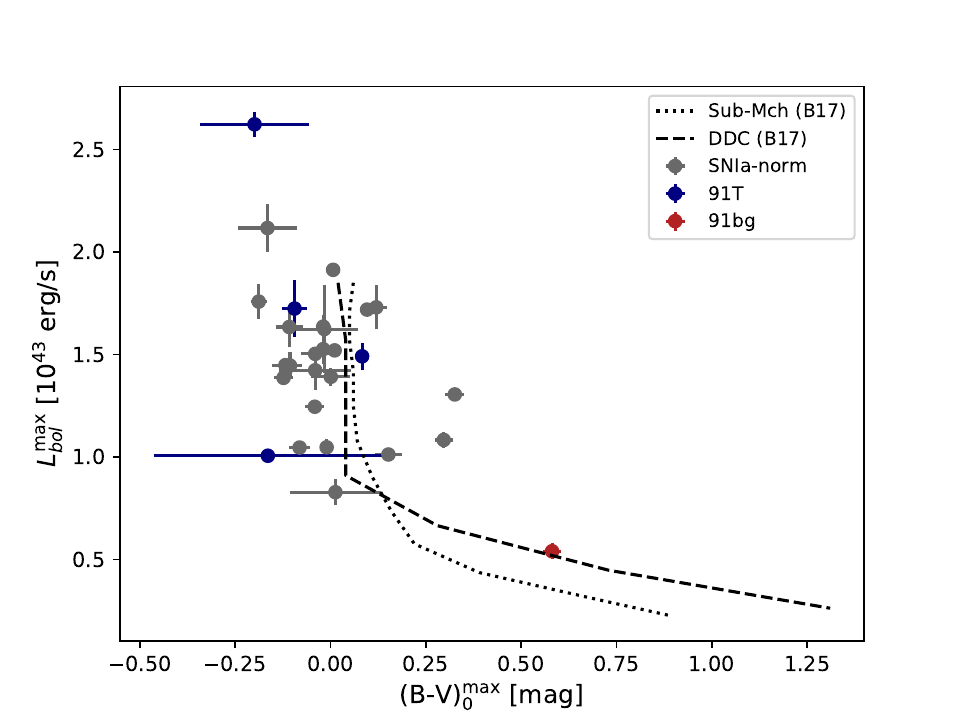}
		\caption{The peak bolometric luminosity against the $(B-V)_0$ color at the moment of maximum. We compare the data with the sub-Chandra detonation (dotted curve) and Chandrasekhar-mass delayed detonation (dashed curve) models of \citet{blondin17}. The blue, grey and red markers show the 91T-, normal- and 91bg subtypes, respectively.} \label{fig:color_lmax_models}
	\end{figure}

	In Figure~\ref{fig:color_lmax_models} we plot the correlation between the inferred bolometric luminosity and the reddening-corrected $(B-V)_0$ color at the moment of maximum light. We compare these data with the results of two explosion models given by \citet{blondin17}: we plot the results of the central detonation sub-Chandra scenario with a dotted curve, while the dashed curve shows the same from the Chandrasekhar-mass delayed-detonation model. It can be seen that the models follow the observed trend of most SNe Ia having $(B-V)_0 \approx 0.0$ mag around maximum, with the fainter events being redder. However, considering just the luminosity and the  color at the moment of maximum is not enough to differentiate between these two explosion scenarios, as they both fit our observed sample relatively well. Our observations also show brighter and bluer events than those predicted by the \citet{blondin17} models.
	
	Overall, we conclude that 
	the predictions of current sub-Chandra models are consistent with the ejecta and nickel masses measured from observations for most SNe~Ia in our sample. Due to the uncertainties of the inferred masses, the conventional delayed-detonation models are also compatible with many of the SNe in our sample, but those models have difficulties explaining the significant number of sub-Chandra ejecta masses.
	
	\subsection{Concerns and caveats regarding the methodology}\label{sec:caveats}
	
	The Arnett-model as well as similar semi-analytic LC models have been widely used for inferring ejecta parameters of various types of SNe, even though such a model is (of course) far from being perfect, and suffers from several issues. Regarding the ejecta mass, one of the most serious, well-known issues is the degeneracy between $M_{\rm ej}$ and the optical opacity $\kappa$, as seen in Equation~(\ref{eq:kappa}): from the $t_{LC}$ timescale, representing the pre-maximum part of the LC, both $\kappa$ and $v_{\rm exp}$ must also be known to calculate $M_{\rm ej}$. $\kappa$, however, is treated very approximately within the framework of the Arnett-model: it is assumed to be constant both in space (within the ejecta) and in time (i.e., no evolution), neither of which is expected to be true in reality. Thus, $\kappa$ should be considered only as a technical, rather than a self-consistent physical parameter, which characterizes the {\it timescale} of the diffusion of photons within a spherical, constant-density, homologously expanding ejecta.

	\begin{figure}[ht]
		\centering
		\includegraphics[width=1\linewidth]{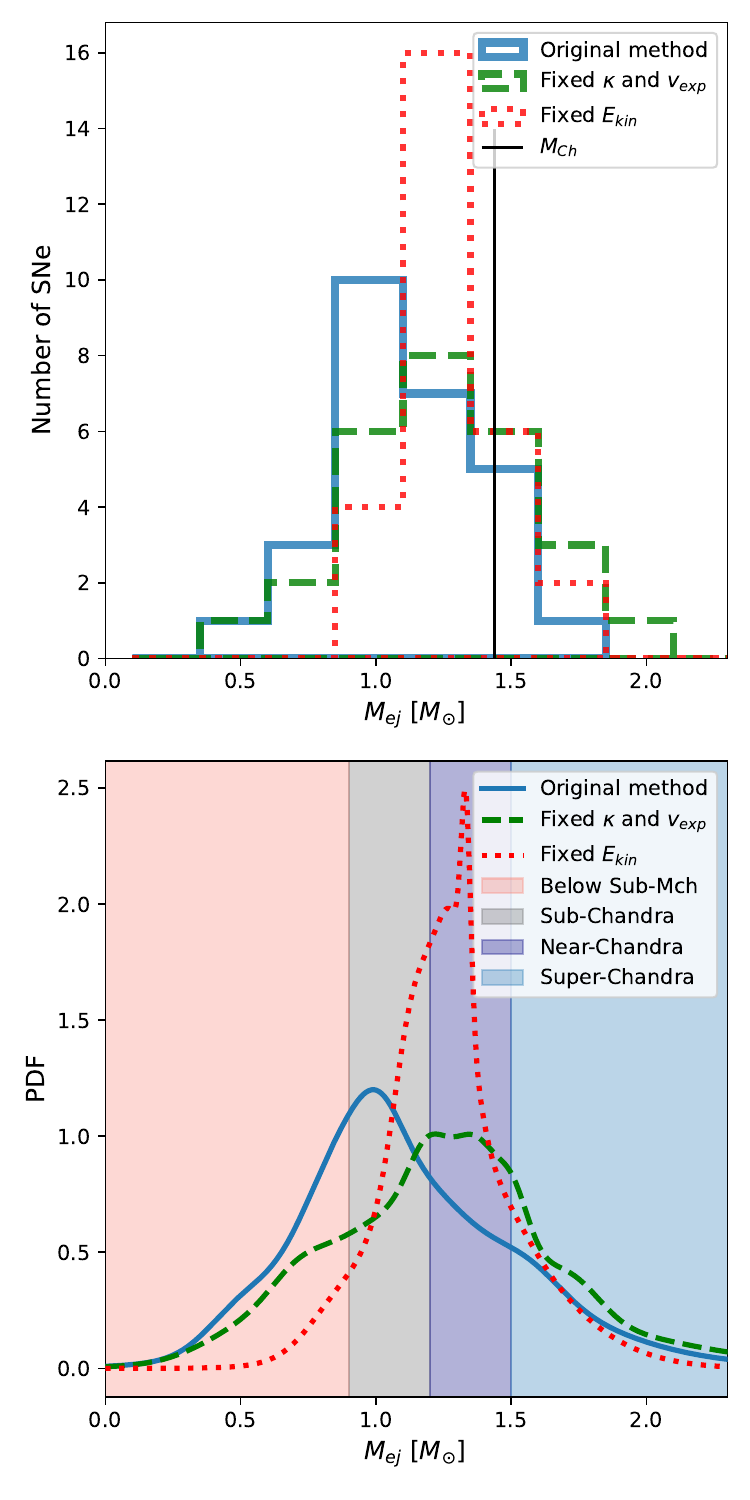}
		\caption{Top panel: Histogram of ejecta masses listed in Table \ref{tab:phys_prop} (blue), masses inferred from Eq.\ref{eq:kappa} assuming uniform opacity (green), and from Eq.\ref{eq:mej} assuming uniform E$_{\rm kin}$ (red). The black vertical line indicates the Chandrasekhar-mass. \\
			Bottom panel: Sum of Gaussian distributions for the same ejecta masses as in the top panel. The background is color-coded by the same mass ranges as in Figure~\ref{fig:ktr_bzs_comp_hist}.}
		\label{fig:mej2}
	\end{figure}
	
	Despite this well-known caveat, it is somewhat surprising that applying physically motivated estimates for $\kappa$, like the expected Thompson-scattering opacity of a metal-dominated plasma ($\kappa \sim 0.1$ cm$^2$g$^{-1}$), one can get quite reasonable ejecta mass estimates for SNe Ia from observationally inferred $t_{LC}$ parameters. 
	In order to verify this statement using the present sample, we explored two alternative ways in deriving the ejecta masses. As a first experiment, we
	calculated ejecta masses from Equation~(\ref{eq:kappa}) assuming $\kappa = 0.1$ cm$^2$g$^{-1}$ and $v_{exp} = 10,000$ km~s$^{-1}$. Second, we substituted $v_{\rm exp}$ into Equation~\ref{eq:mej} with the kinetic energy $E_{\rm kin}$ taken from Equation~\ref{eq:ekin}, expressed $M_{\rm ej}$, and assumed $E_{\rm kin} = 10^{51}$ erg as a uniform value for the whole sample.  
	The results of these experiments are plotted as histograms and Gaussian PDFs in Figure~\ref{fig:mej2}, together with the distribution of the original $M_{\rm ej}$ values listed in Table~\ref{tab:phys_prop}. 
	
	It is seen that, despite the over-simplifying assumptions used in these experiments, either the constant opacity and expansion velocity, or the kinetic energy that are uniform for the whole Ia sample, the resulting ejecta masses are close to the ones estimated from the gamma-ray leaking timescales, even though the peaks of their distributions are shifted to somewhat higher values. The mean ejecta mass from the uniform opacity assumption is $\langle M_{\rm ej} \rangle = 1.32$ M$_\odot$ with a standard deviation of $0.47$ M$_\odot$, while from the uniform kinetic energy assumption these are $\langle M_{\rm ej} \rangle = 1.26$ M$_\odot$ and $0.18$ M$_\odot$. It is seen that using such kind of physically-motivated assumptions one can get slightly different ejecta masses than by using the data-driven method we preferred in this paper, which makes as minimal assumptions as possible within the framework of the Arnett-model. Still, even if we adopted any of the results from the above experiments, a non-negligible fraction of the sample
	($\sim 7$\% in the uniform opacity model, or $\sim 19$\% in the uniform kinetic energy model) would be consistent only with sub-Chandra explosions (see Table~\ref{tab:statistics_yeah}). 
	However, the uniform opacity model would also predict that
	$\sim 26$\% of the sample is consistent only with near-Chandra explosions, with $\sim 22$\% of the sample being super-Chandra ($M_{\rm ej} > 1.5$ M$_\odot$).  In the fixed kinetic energy model the fraction of only near-Chandra explosions is higher ($\sim 30$\%). The number of SNe that are consistent with both explosion channels are nearly the same (40 -- 50\%), regardless of the adopted LC interpretation. Thus, it seems that the Arnett-models that relate the {\it timescales} of the SN bolometric light curves to the basic physical parameters of the ejecta are consistent with the observations only if a fraction of the observed SNe Ia are produced by sub-Chandra progenitors. 
	
	While discussing the physics of SNe Ia, \citet{lm18} argued that ejecta mass estimates based on the Arnett-model are sensitive only to the opaque mass. Since the optical opacity is mostly sensitive to the Fe-group elements, such a method can recover only the mass rich of Fe-group elements. They concluded that {\it ``It would therefore not be surprising for Scalzo et al. to find masses that are proportional to the nickel mass''.}  This argument, however, assumes that the ejecta mass estimate is based on the optical (average) opacity, like in Equation~(\ref{eq:kappa}). It is emphasized that the ejecta mass estimates that are based on the gamma-ray transparency timescale ($t_\gamma$) that we use in this paper (and also the numerous previous studies discussed in Section~\ref{sec:discussion}), is much less affected by the uncertainty of the gamma-ray opacity, $\kappa_\gamma$, which is constrained much better and it is less sensitive to the actual chemical composition of the ejecta than the optical opacity. 
	
	In this study we have adopted $\kappa_\gamma = 0.025$ cm$^2$~g$^{-1}$ from \citet{guttman24}, who derived the value from a semi-analytic method. Some previous works used $\kappa_\gamma = 0.03$ cm$^2$~g$^{-1}$ from \citet{wheeler15} but this value goes back to \citet{colgate80} who inferred $\kappa_\gamma = 0.028$ cm$^2$~g$^{-1}$ using a multiple-scattering Monte Carlo gamma-ray transport code developed at Los Alamos. It is seen that these values are in good agreement with each other. Still, because $\kappa_\gamma$ appears in the denominator in Equation~(\ref{eq:mej}), the inferred $M_{\rm ej}$ masses are sensitive to the adopted value of this parameter. In order to test the effect of the adopted gamma-ray opacity on the final results, we have re-calculated the ejecta masses from Equation~(\ref{eq:mej}) using $\kappa_\gamma = 0.028$ and 0.03 cm$^2$~g$^{-1}$.
	
	\begin{figure}[ht]
		\centering
		\includegraphics[width=1\linewidth]{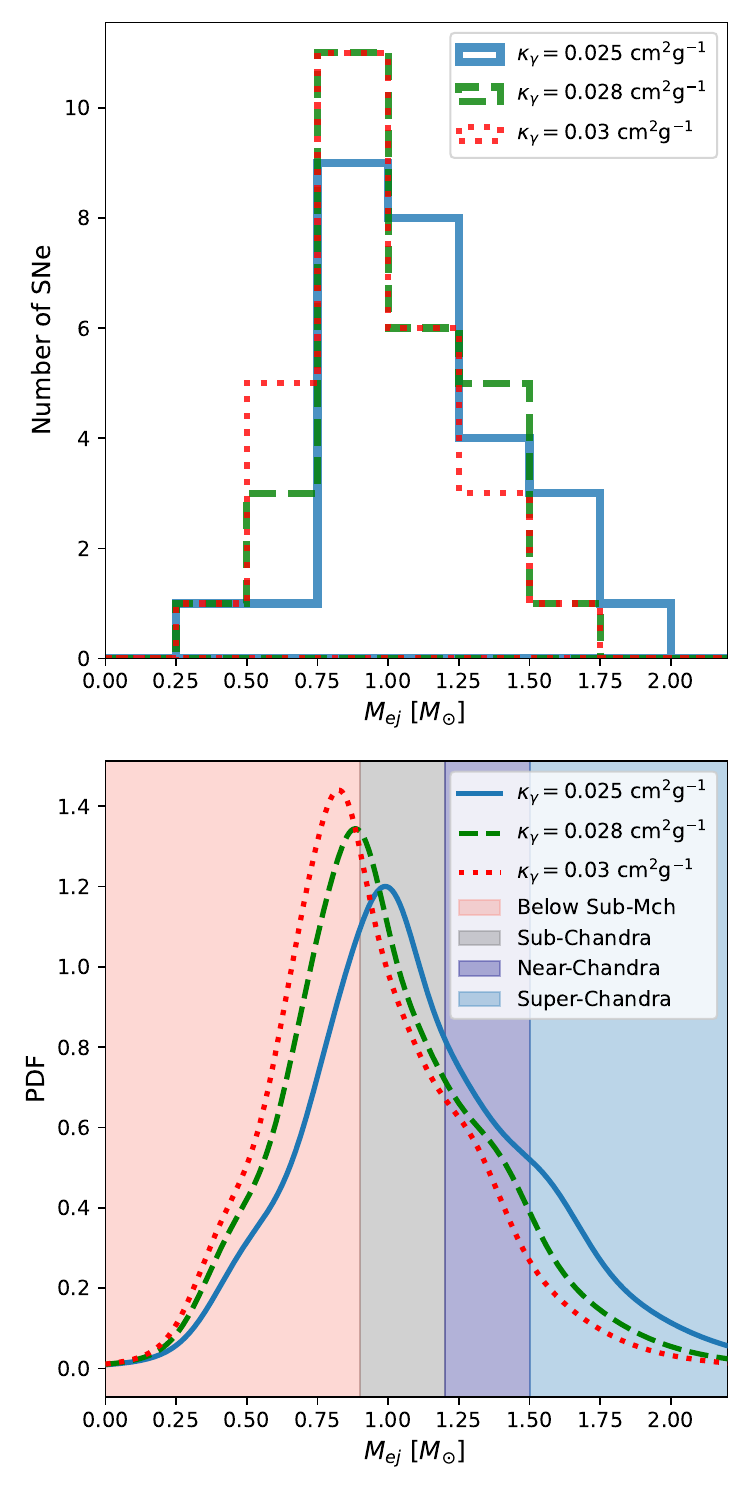}
		\caption{Top panel: Histograms of the ejecta masses inferred from Eq.\ref{eq:mej} assuming different $\kappa_{\gamma}$ values. We show the ejecta masses obtained with $\kappa_\gamma = 0.025$ cm$^2$~g$^{-1}$ (solid blue), $\kappa_\gamma = 0.028$ cm$^2$~g$^{-1}$ (dashed green), and $\kappa_\gamma = 0.03$ cm$^2$~g$^{-1}$ (dotted red).\\
			Bottom panel: Gaussian distributions of the different masses, constructed the same way as in Figure~\ref{fig:ktr_bzs_comp_hist}.}
		\label{fig:gamma_histograms}
	\end{figure}
	
	\begin{figure}
		\centering
		\includegraphics[width=1\linewidth]{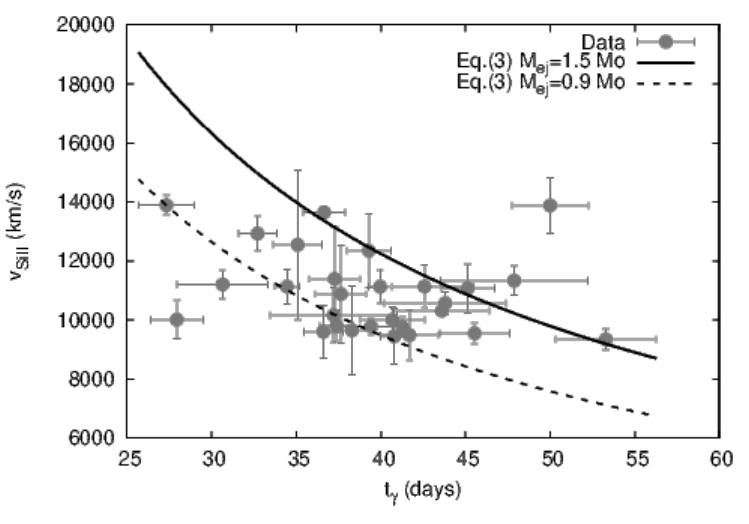}
		\caption{Observed $v_{\rm SiII}$ velocities (filled symbols) compared with the expansion velocities for fixed ejecta masses against the gamma-ray leaking timescale $t_\gamma$. The fixed ejecta masses corresponding to the continuous and dashed curves are shown in the legend. }
		\label{fig:vel-vel}
	\end{figure}

	The results are shown in Figure~\ref{fig:gamma_histograms}, where the histograms and PDFs of the inferred $M_{\rm ej}$ values corresponding to the different adopted gamma-ray opacities are plotted together. It is seen that adopting a slightly higher gamma-ray opacity decreases the inferred ejecta masses (Table \ref{tab:statistics_yeah}), as expected. This can be considered as a systematic uncertainty of the calculated ejecta masses. However, the range of the potential gamma-ray opacity values is small enough that this systematic uncertainty is not severe. Indeed, as Table~\ref{tab:statistics_yeah} shows, adopting $\kappa_\gamma = 0.03$~cm$^2$~g$^{-1}$ would decrease only the number of systems that are consistent only with the near-Chandra scenario, but the fraction of sub-Chandra SNe would be nearly the same. It is true, however, that the higher gamma-ray opacity would significantly increase the fraction of SNe that have too low ejecta masses, hence they would become inconsistent even with the sub-Chandra explosion models. Still, regardless of the adopted gamma-ray opacity value, the majority of the SNe in our sample are consistent with both explosion channels.

	These calculations confirm that the ejecta masses inferred from the gamma-ray transparency timescales measured at late phases suggest that more SNe Ia may be due to sub-Chandra explosions than what could be estimated from the near-maximum light curves. The key parameter behind this conclusion is clearly the gamma-ray opacity, which is much better constrained than the optical (electron scattering) opacity.

	\begin{table*}[ht]
		\centering
		\caption{Comparison of the sample SNe with explosion models based on the inferred ejecta masses}
		\begin{tabular}{l|c|c|c|c|c|c}
			\hline
			Method  & $\langle M_{\rm ej} \rangle$  &   $\sigma(M_{\rm ej})$ &  Consistent only &   Consistent only  &   Consistent    &   Outside of \\
			&                  &   & with Sub-M$_{\rm Ch}$ &  with near-M$_{\rm Ch}$  &   with both  &    both channels\\
			\hline
			Fixed $\kappa$, v$_{\rm exp}$               & 1.32    &   0.47  &   7\% &   26\%    &   44\%   &    22\% \\
			Fixed E$_{\rm kin}$                         & 1.26    &   0.18 &   19\%    &   30\%    &   52\%  & 0\%  \\
			$\kappa_{\gamma}$ = 0.025 cm$^2$~g$^{-1}$   & 1.12    &   0.30 &   33\%    &   15\%  &  48\%  & 4\%  \\
			$\kappa_{\gamma}$ = 0.028 cm$^2$~g$^{-1}$   &  0.99   &   0.27   &   37\%    &   4\%    &   44\%     &   15\%   \\
			$\kappa_{\gamma}$ = 0.030 cm$^2$~g$^{-1}$    & 0.93    &   0.25  &   33\%    &   0\% &   41\%  & 26\%  \\
			
			\hline
		\end{tabular}
		\tablenotetext{}{First column: method for calculating $M_{\rm ej}$ . 2nd and 3rd  columns: the mean and the standard deviation of the masses inferred by each method. Next columns: the percentage of SNe 
			that are consistent with only sub-Chandra, only near-Chandra, both or neither explosion models. See Section~\ref{sec:caveats} for explanation.}
		\label{tab:statistics_yeah}
	\end{table*}

	The expansion velocity parameter, $v_{\rm exp}$, is also problematic. In the original paper \citet{arnett82} used the term ``scaling velocity'', $v_{\rm sc}$, to emphasize that it is more like a parameter describing the homologous expansion of a constant-density sphere instead of being an actually observed velocity. However, he also argued that {\it ``... if we take $v_{\rm sc}$ from the photospheric fluid velocity near maximum light, this value is not far from the uniform-density estimate''.} This note was the main motivation for us to assign the observed $v_{\rm SiII}$ at maximum to the expansion velocity in our calculations. 
	
	In order to test whether the observed $v_{\rm SiII}$ velocities were valid expansion velocities in the Arnett-model, we inferred the expected expansion velocity as a function of $t_{\gamma}$ from Equation~(\ref{eq:mej}) by assuming a fixed ejecta mass, and compared them with the observed velocities (see Figure~\ref{fig:vel-vel}). It is seen that most of the observed points fall within the two theoretical curves corresponding to $M_{\rm ej} = 0.9$ and 1.5~M$_\odot$. Thus, using the observed $v_{\rm SiII}$ velocity and $t_\gamma$ values one can get reasonable estimates for the ejecta mass via Equation~(\ref{eq:mej}).

	Still, the question remains whether $v_{\rm SiII}$ gave the {\it total} ejecta mass when inserting it into Equation~(\ref{eq:mej}). In other words: is it possible that $v_{\rm SiII}$ probes only the inner, optically thick part of the ejecta at the moment of maximum light, resulting in a mass that is systematically lower than the true $M_{\rm ej}$? This is a non-trivial problem, because the Arnett-model assumes a constant density ejecta, while real SNe Ia have different density profiles. The detailed investigation of this issue will be the subject of a subsequent paper (Vinkó et al., in preparation), but a quick analytic estimate is given here. 
	Following \citet{magee20}, we used the exponential density profile of a fiducial SN Ia having $M_{\rm ej} = 1.44$~M$_\odot$ and $E_{\rm kin} = 10^{51}$ erg to calculate the velocity at the photosphere via the usual condition of $\tau(v_{\rm ph}) = 2/3$, where $\tau(v_{\rm ph})$ is the optical depth of the ejecta at $v_{\rm ph}$. After that, the mass above $v_{\rm ph}$, which is optically thin and may not contribute to the photon diffusion, was determined by integrating the density profile between $v_{\rm ph}$ and $v_{\rm max} = 30,000$~km~s$^{-1}$ \citep{magee20}. Adopting $t_{\rm max} \approx 18$ days as the moment of maximum light after explosion, we got $v_{\rm ph} = 11,571$~km~s$^{-1}$ and $13,238$~km~s$^{-1}$ for $\kappa = 0.1$ and 0.2 cm$^2$~g$^{-1}$, respectively. 
	The masses above these photospheric velocities are $\Delta M_{\rm ej} \approx 0.20$ and 0.13~M$_\odot$, respectively. It is seen that even though the mass probed by the Arnett-model is $1.44 - 0.20 = 1.24$ ~M$_\odot$, i.e., $\sim 14$\% lower than $M_{\rm Ch}$, it would still be within the range of the near-Chandra explosions. Thus, it is not likely that the significant fraction of the sub-Chandra ejecta masses given by the Arnett-model are simply due to the systematically low velocities inserted into Equation~(\ref{eq:mej}).

	Yet another important restriction of the Arnett-model is the assumption of a centrally localized heating source, i.e., the spatial distribution of $^{56}$Ni is strongly peaked at the center. This may be a fair assumption for a fiducial SN Ia model having $M_{\rm Ch}$ mass and $\sim 0.6$~M$_\odot$ (i.e., $\sim 40$\% mass fraction) of $^{56}$Ni. However, it can be expected to break down in several cases: $i)$ when $M_{\rm Ni} \gtrsim 1.0$ M$_\odot$, i.e., the majority of the ejecta mass consists of radioactive Ni, $ii)$ when $M_{\rm ej} << M_{\rm Ch}$, i.e., for sub-Chandra explosions, and $iii)$ when $^{56}$Ni has a more extended distribution for other reasons, e.g., off-center ignition \citep{seitan13, magee20}.  If the radioactive heating is not centrally localized, the photon diffusion timescale may get shorter, and some of the gamma-rays produced by the radioactive decay may start to leak out from the expanding ejecta earlier than expected for a centrally located heating source. These effects may result in shorter $t_{\rm LC}$ and $t_{\gamma}$ timescales for the observed light curve, which, in the formalism of the Arnett-model, gives an ejecta mass that is lower than in reality.
	
	This caveat may explain the presence of SNe having ejecta masses significantly less than the lower limit of sub-Chandra Ia models (0.9~M$_\odot$), which otherwise have ``normal'' nickel masses in the observed sample (Figure~\ref{fig:mni_mej_models}). This may also be the explanation for those objects having $M_{\rm Ni} \gtrsim M_{\rm ej}$ (SN~2021dov or SN~2017erp in KTR20). Since it is certainly a limitation for the applicability of the Arnett-model to real data, the resulting too low ejecta masses or the cases when $M_{\rm Ni} / M_{\rm ej} \approx 1$ should be treated with caution.

	Even though these undeniable caveats exist, we believe that radiation-diffusion (Arnett-) models can still give us reliable constraints on the global physical properties of SNe Ia if the photometry extends well beyond maximum light into the regime when the ejecta becomes semi-transparent to gamma-rays, and there is near-maximum spectroscopy available.

	\section{Summary}   \label{sec:summa}
	We presented multicolor photometry of 28 nearby Type Ia supernovae taken during the first 4 years of operation of the RC80 and BRC80 telescopes at Piszkéstető station of Konkoly Observatory and Baja Observatory of University of Szeged, Hungary. Using the multicolor light curve fitting codes SALT3 and MLCS2k2, we determined luminosity distances as well as total dust extinctions to the sample SNe, which we used to construct their pseudo-bolometric light curves. Fitting the bolometric light curves with the Arnett-model yielded the mass of the radioactive nickel synthesized during the explosion along with the important timescales of $t_{\mathrm{LC}}$, and $t_{\gamma}$, which can be used to estimate the ejecta mass if the photospheric velocity is known. After collecting previously unpublished optical spectra taken by LCO telescopes and the Hobby-Eberly Telescope, supplemented by public spectra downloaded from the Transient Name Server, we measured the velocities of the Si~II $\lambda$6355 feature and determined the $v_{\rm SiII}$ velocities at the moment of maximum for all of our sample SNe. 
	
	Via Equations~(\ref{eq:mej}), (\ref{eq:kappa}) and (\ref{eq:ekin}) we calculated the ejecta masses, optical opacities and kinetic energies for the sample SNe. 
	We found that based on 
	the inferred ejecta masses, $\sim 48$\% of the sample SNe are consistent with both sub-Chandra and near-Chandra explosion models. 
	Our results suggest that $\sim 33$\% of the sample can be explained by only sub-Chandra progenitors. This is a somewhat lower value than appeared earlier in the literature \citep{scalzo14a, scalzo14b, scalzo19, ktr20, liu23_2}, but together with the events that are consistent with both explosion channels, this could be as high as 81\%.  The fraction of near-Chandra events was found to be in between $\sim 15$--63\%, but this estimate may suffer from low-number statistics and can also be affected by the adopted value of the gamma-ray opacity. If we used $\kappa_\gamma = 0.03$ cm$^2$~g$^{-1}$ instead of 0.025, the maximum fraction of near-Chandra SNe Ia would decrease down from 63 \% to $\sim 41$\% at best.
	
	After comparing our results with multiple types of recent explosion models, we revealed that significant differences between the predictions of near-Chandra and sub-Chandra models cannot be seen in the color-luminosity relation around maximum (see Figure~\ref{fig:color_lmax_models}). We showed that the ejecta mass can be a useful constraint to confront the predictions of various explosion models with the observations. From the results presented in this paper it seems that the majority of the studied SNe Ia are consistent with either the sub-Chandra or the near-Chandra explosion models, but the accuracy of the current photometric mass estimates are not high enough to clearly distinguish between these two possibilities.

	\begin{acknowledgments}
		We thank the anonymous referee for their helpful comments and guidance in improving our paper.
		
		This research is supported by  NKFIH-OTKA grants K-142534, K-134432, K-138962, FK-134432, and PD-134784, NKFIH \'Elvonal grant KKP-143986, and  NKFIH excellence grant TKP2021-NKTA-64 from the National Research, Development and Innovation Office (NKFIH), Hungary. The RC80 and BRC80 telescopes have been supported by the GINOP 2.3.2-15-2016-00033 project from the Government of Hungary, funded by the European Union. JCW and JV are supported by NSF grant AST-1813825. \'AS, CK, and AB acknowledge  financial support of the KKP-137523 'SeismoLab' \'Elvonal grant of NKFIH, Hungary. 
		BC received support from the Lend\"ulet Program LP2023-10 of the Hungarian Academy of Sciences. KV and LK are supported by the Bolyai J\'anos Research Scholarship of the Hungarian Academy of Sciences. Financial support from the Austrian--Hungarian Action Foundation grants 112\"ou1 is also acknowledged. 
		ZB is supported by the ÚNKP-22-2 New National Excellence Program of the Ministry for Culture and Innovation from the source of the National Research, Development and Innovation Fund.
		VV is supported by the ÚNKP-22-1 New National Excellence Program of the Ministry for Culture and Innovation from the source of the National Research, Development and Innovation Fund.
		ZB, ÁH-D, NOS, and VV thank the financial support provided by the undergraduate research assistant program of Konkoly Observatory.
		LK acknowledges the Hungarian National Research, Development and
		Innovation Office grant OTKA PD-134784.
		Zs.M.Sz. acknowledges funding from a St Leonards scholarship from the University of St Andrews, and is a member of the International Max Planck Research School (IMPRS) for Astronomy and Astrophysics at the Universities of Bonn and Cologne.
		KAB is supported by an LSST-DA Catalyst Fellowship; this publication was thus made possible through the support of grant 62192 from the John Templeton Foundation to LSST-DA. 
		AVF is grateful for financial assistance from the Christopher R. Redlich Fund and many other donors.
		
		This work makes use of data from the Las Cumbres Observatory global telescope network. The LCO group is supported by NSF grants AST-1911151 and AST-1911225.
	\end{acknowledgments}  
	\begin{acknowledgments}  
		
		The Hobby–Eberly Telescope (HET) is a joint project of the University of Texas at Austin, the Pennsylvania State University, Ludwig-Maximilians-Universitat Munchen, and
		Georg-August-Universitat Gottingen. 
		The HET is named in honor of its principal benefactors, William P. Hobby and Robert E. Eberly. The Low Resolution Spectrograph 2 (LRS2) was developed and funded by the University of Texas at Austin McDonald Observatory and Department of Astronomy, and by Pennsylvania State University. We thank the Leibniz-Institut fur Astrophysik Potsdam (AIP) and the Institut fur Astrophysik Goettingen (IAG) for their contributions to the construction of the integral field units. The authors are grateful to the HET Resident Astronomers and staff members at McDonald Observatory and Las Cumbres Observatory for their excellent work.
	\end{acknowledgments}

	\vspace{5mm}

	\facilities{RC80 (Konkoly); BRC80 (U Szeged); HET (McDonald); Las Cumbres Observatory telescopes  }
	
	\software{SALT3 \citep{taylor23}; 
		MLCS2k2 \citep{jha07}; sncosmo \citep{barbary16}; 
		Minim \citep{manos13}}

	\bibliographystyle{aasjournal}
	{}
	
	\appendix
	
	In Figures~16-19 we present plots of the bolometric light curves together with their best-fitting Arnett-models. 
	
	More plots and tables can be found online in the GitHub repository of this paper: \\
	{\tt \url{https://github.com/borazso/PiszkeSN}}.

	\begin{figure*}[h!]
		\centering
		\includegraphics[width=0.40\linewidth]{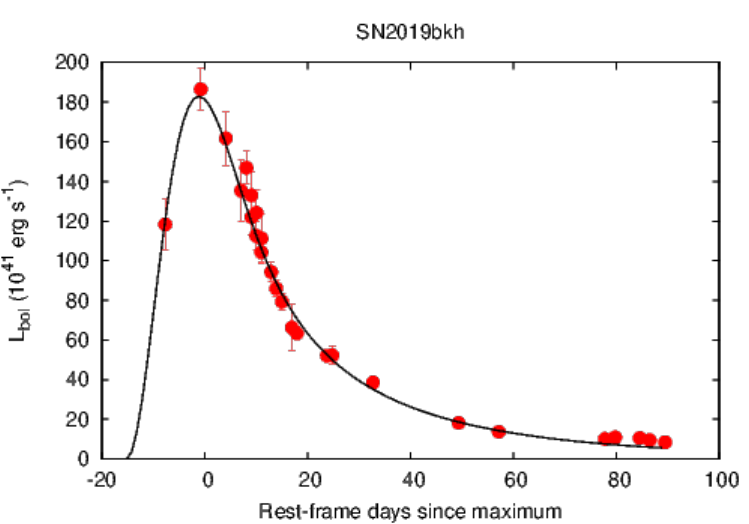}
		\includegraphics[width=0.40\linewidth]{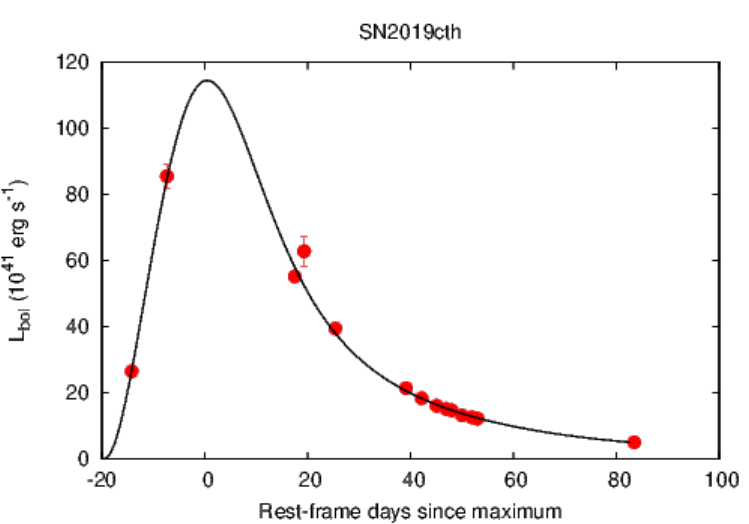}
		\includegraphics[width=0.40\linewidth]{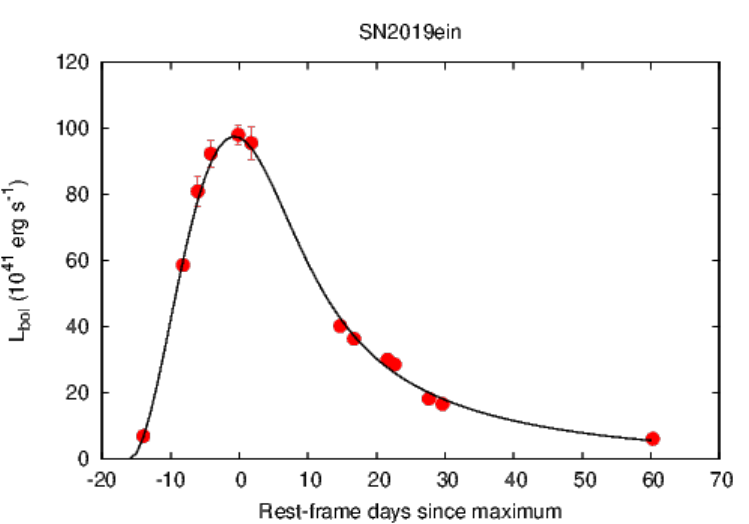}
		\includegraphics[width=0.40\linewidth]{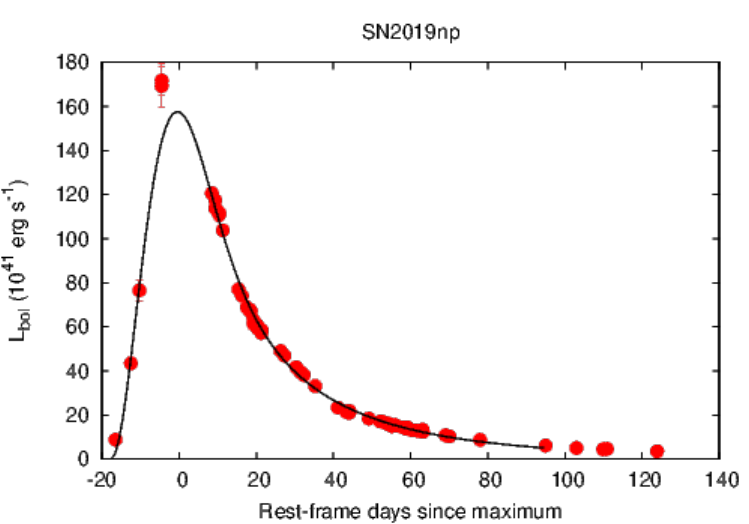}
		\includegraphics[width=0.40\linewidth]{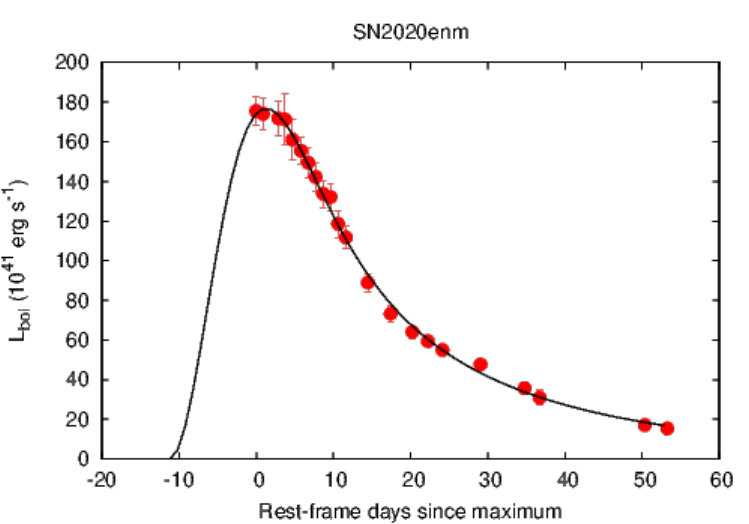}
		\includegraphics[width=0.40\linewidth]{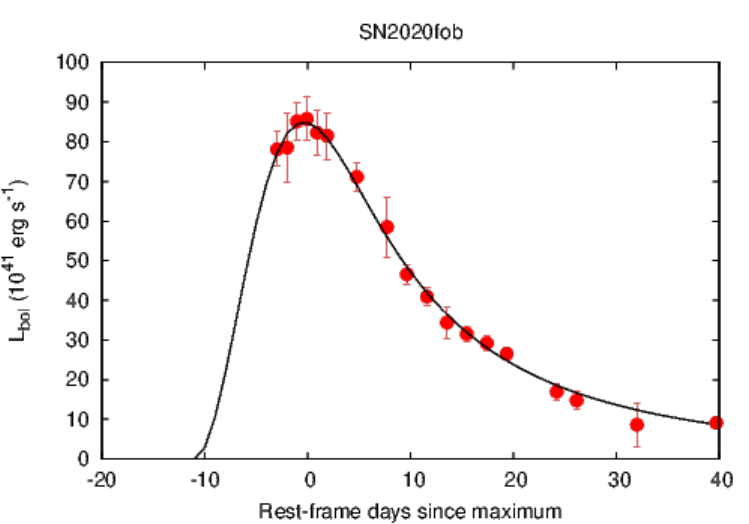}
		
		\caption{Bolometric light curves of the sample SNe (filled symbols) and their best-fitting Arnett-models (black curves).}
		\label{fig:minimplots}
	\end{figure*}
	
	\begin{figure*}[ht]
		\centering
		\includegraphics[width=0.40\linewidth]{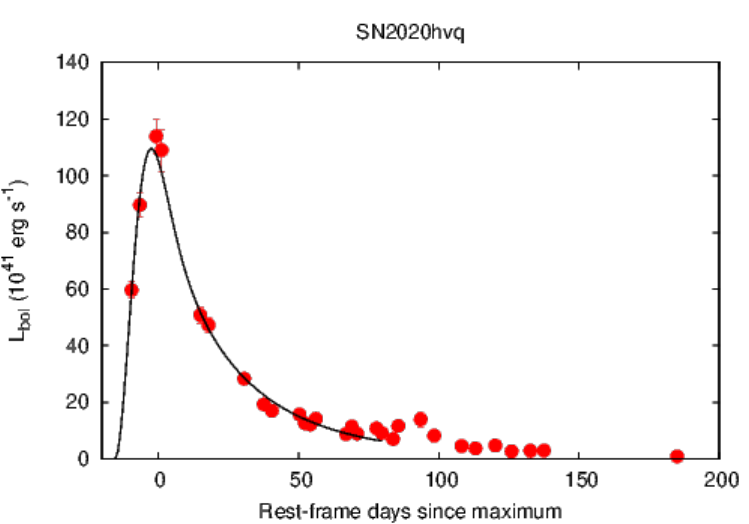}
		\includegraphics[width=0.40\linewidth]{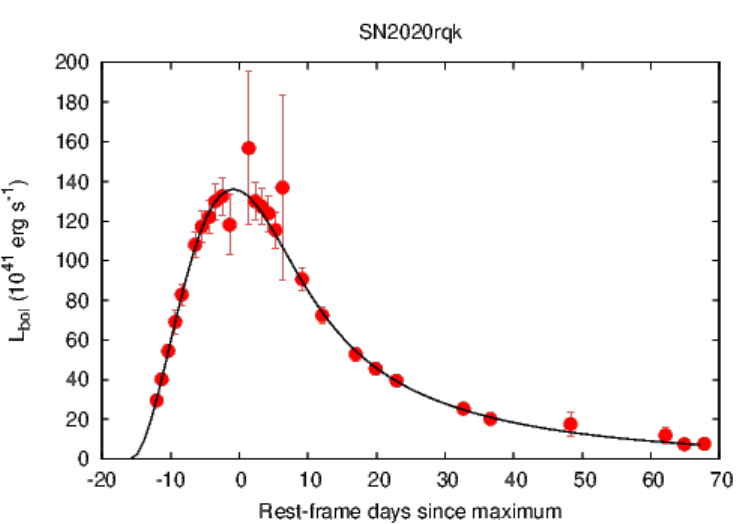}
		\includegraphics[width=0.40\linewidth]{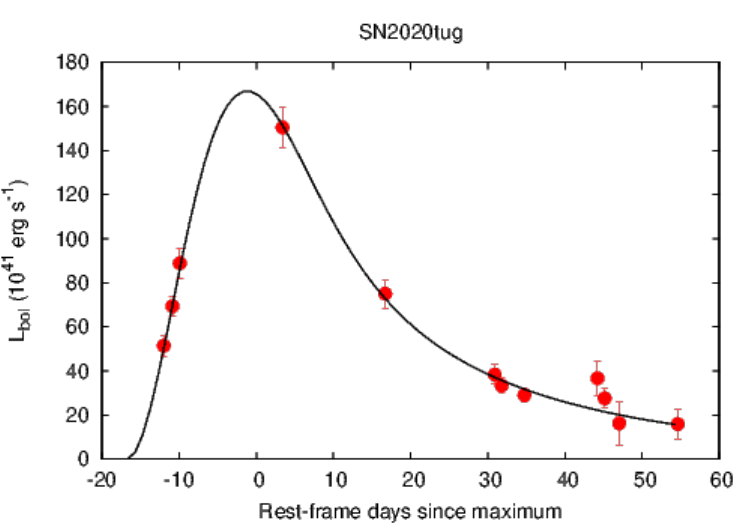}
		\includegraphics[width=0.40\linewidth]{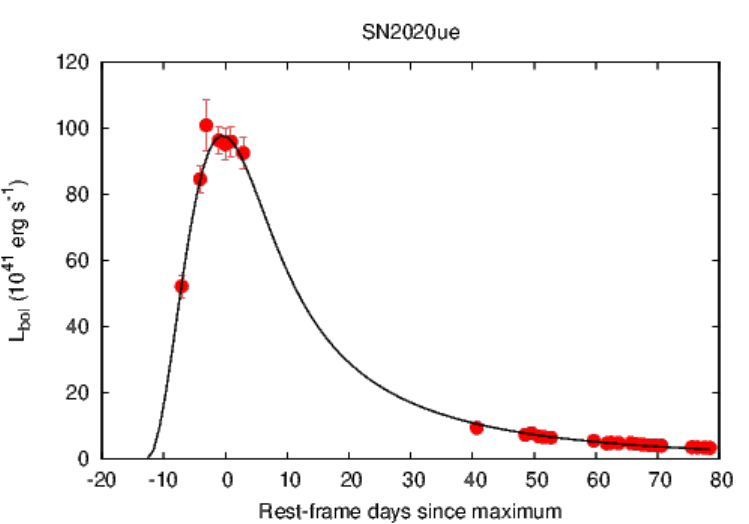}
		\includegraphics[width=0.40\linewidth]{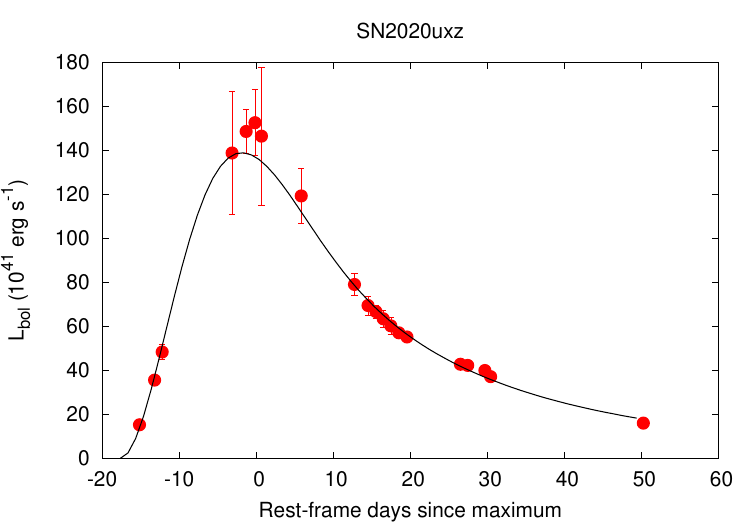}
		\includegraphics[width=0.40\linewidth]{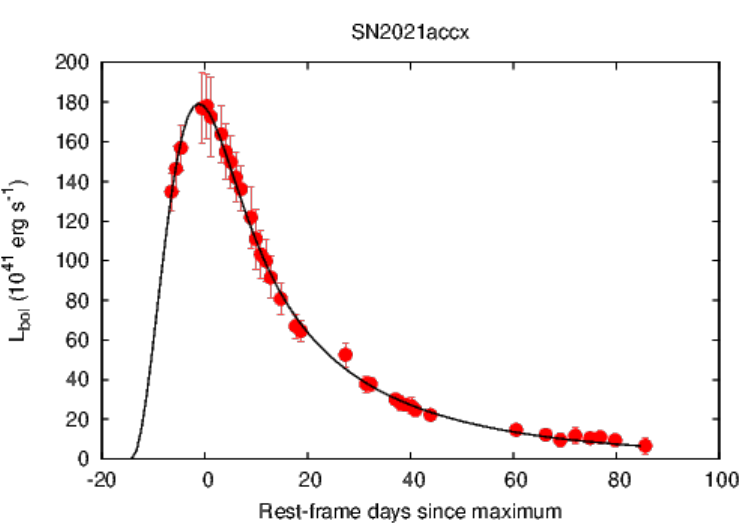}
		\includegraphics[width=0.40\linewidth]{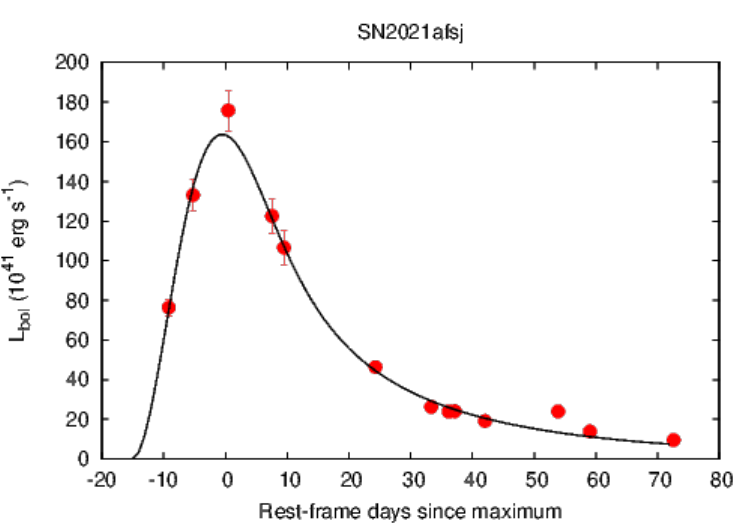}
		\includegraphics[width=0.40\linewidth]{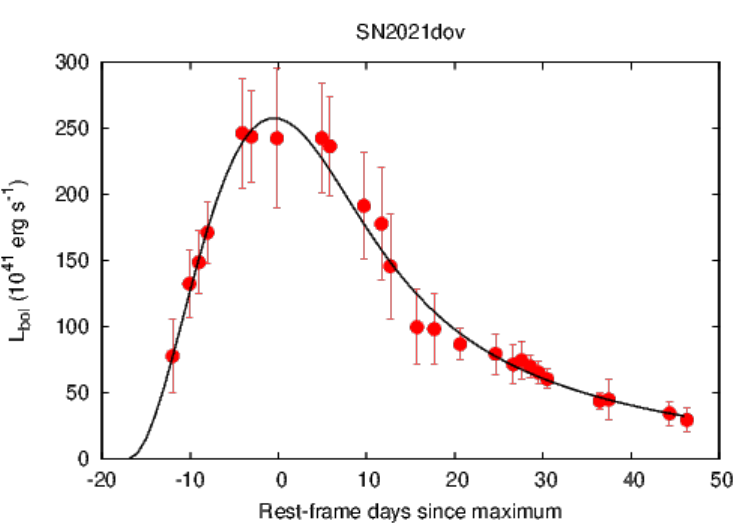}
		\caption{The same as Figure~\ref{fig:minimplots} but for additional SNe.}
	\end{figure*}
	
	\begin{figure*}[ht]
		\centering
		\includegraphics[width=0.40\linewidth]{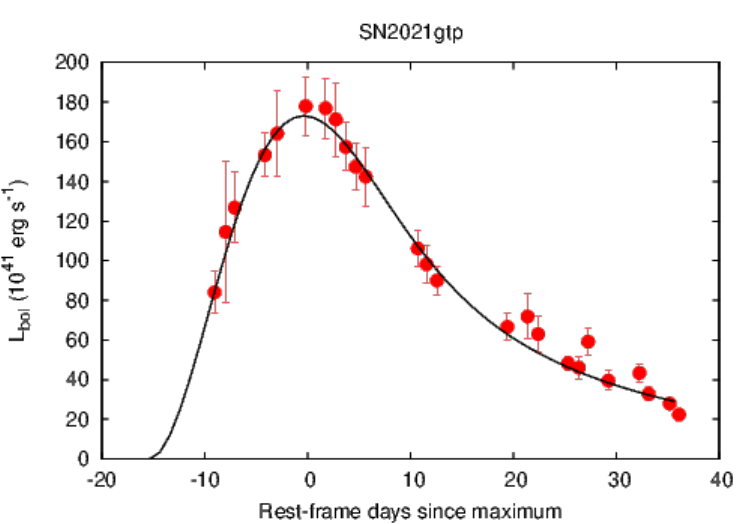}
		\includegraphics[width=0.40\linewidth]{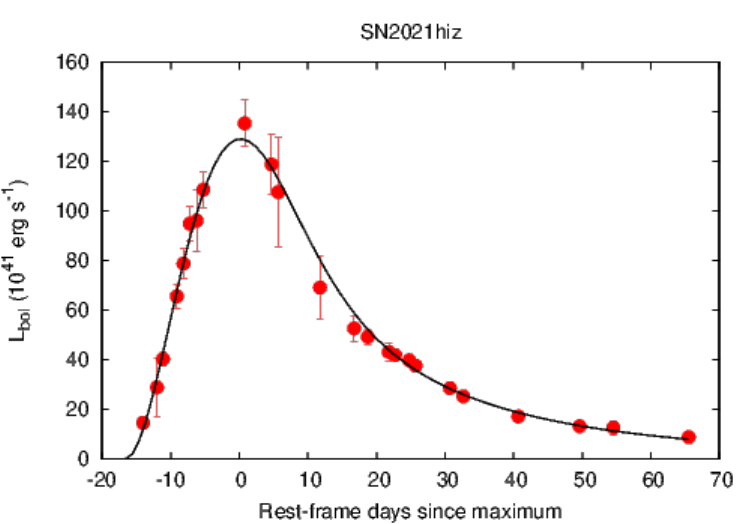}
		\includegraphics[width=0.40\linewidth]{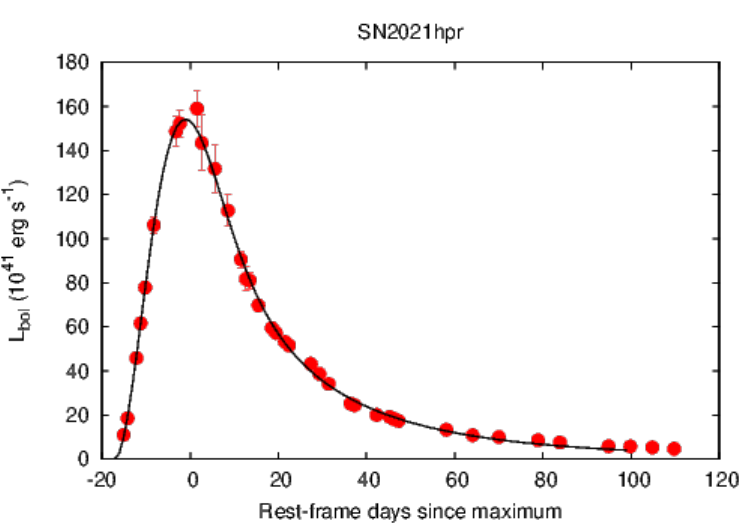}
		\includegraphics[width=0.40\linewidth]{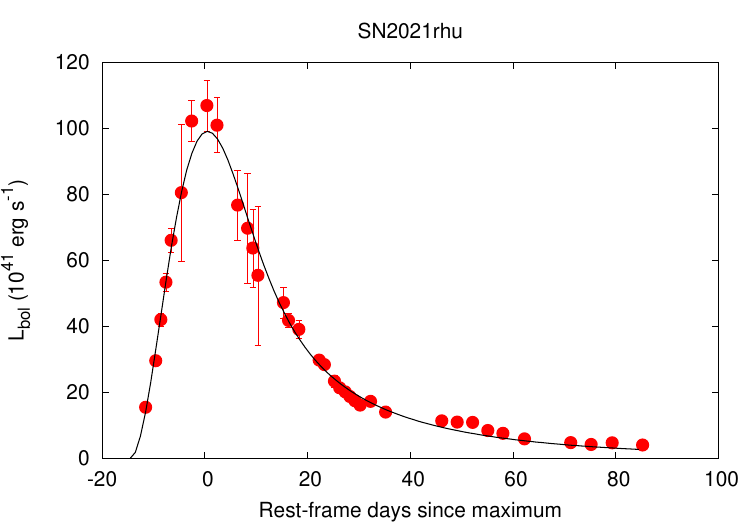}
		\includegraphics[width=0.40\linewidth]{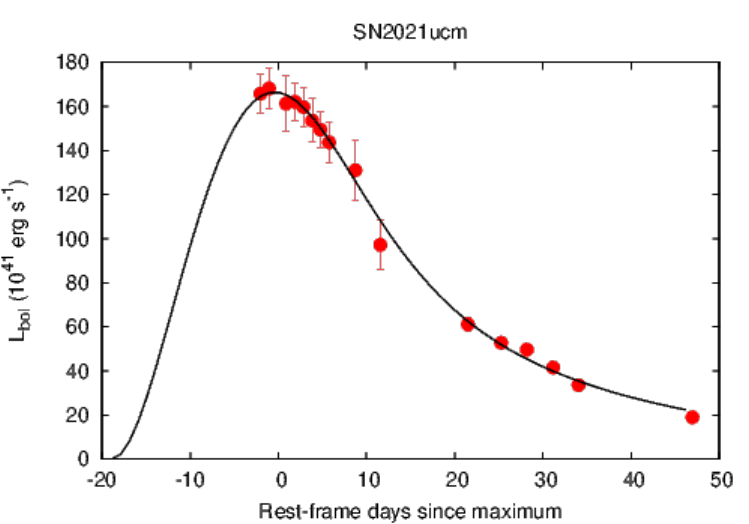}
		\includegraphics[width=0.40\linewidth]{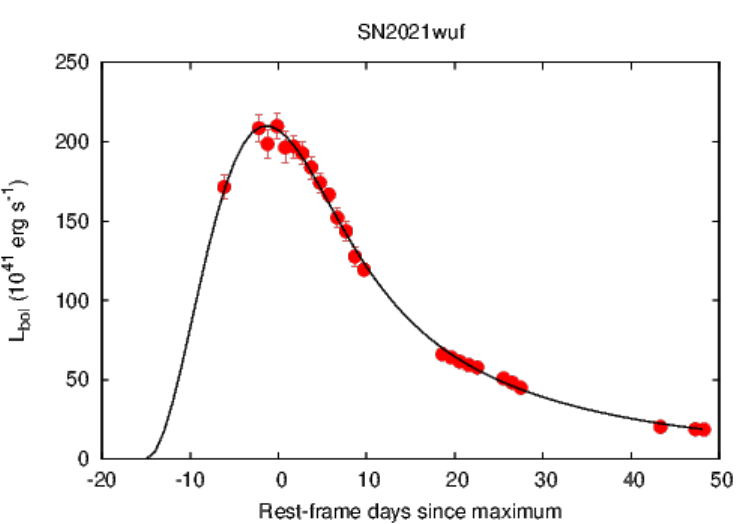}
		\includegraphics[width=0.40\linewidth]{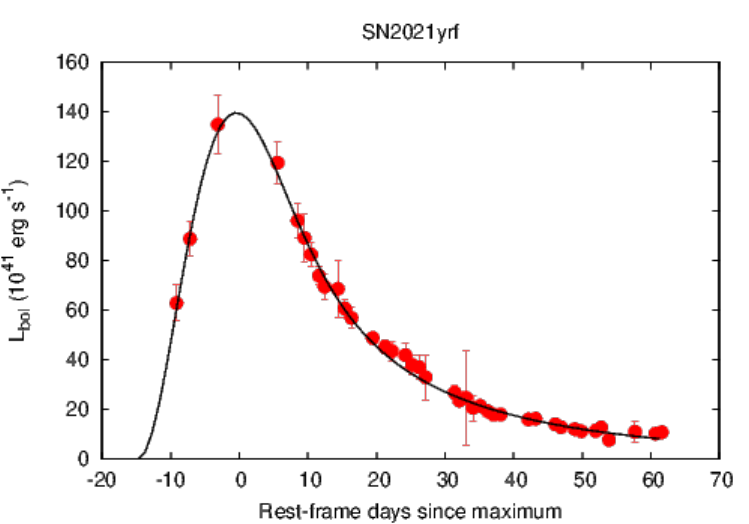}
		\includegraphics[width=0.40\linewidth]{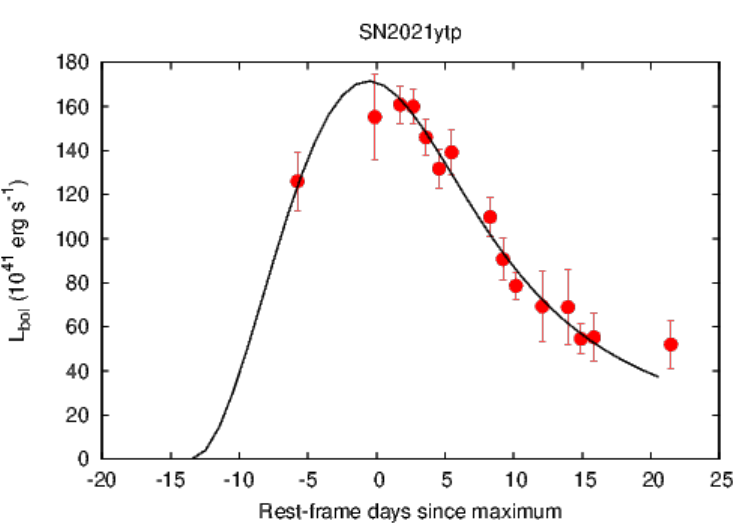}
		\caption{The same as Figure~\ref{fig:minimplots} but for additional SNe.}
	\end{figure*}
	
	\begin{figure*}[ht]
		\centering
		\includegraphics[width=0.40\linewidth]{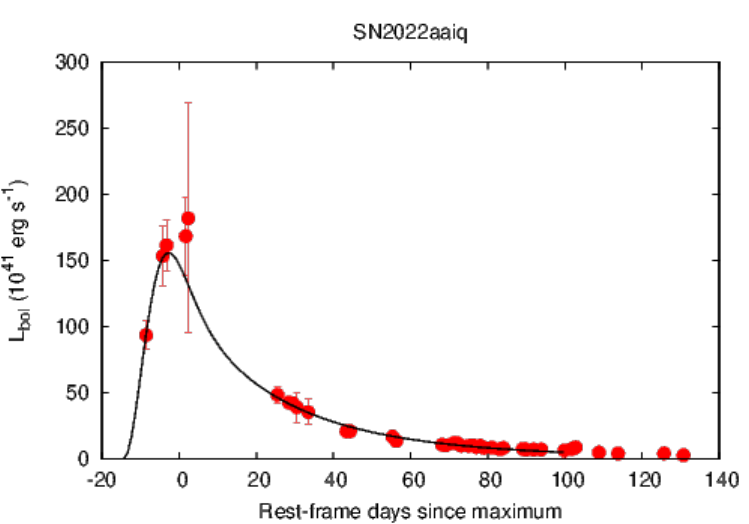}
		\includegraphics[width=0.40\linewidth]{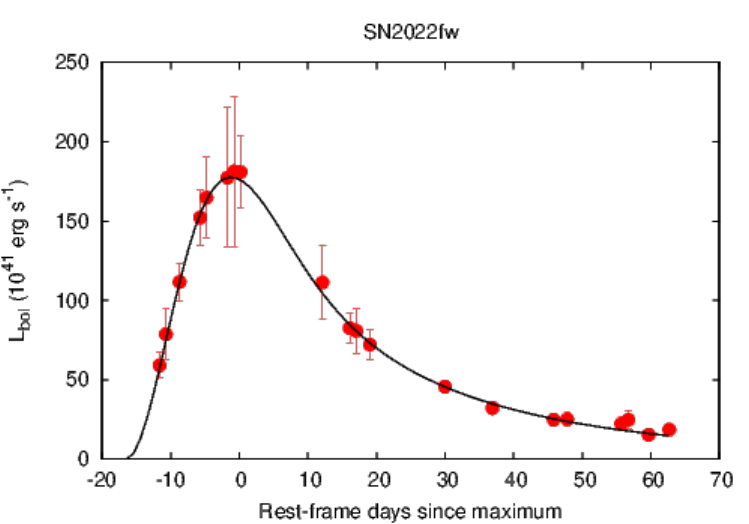}
		\includegraphics[width=0.40\linewidth]{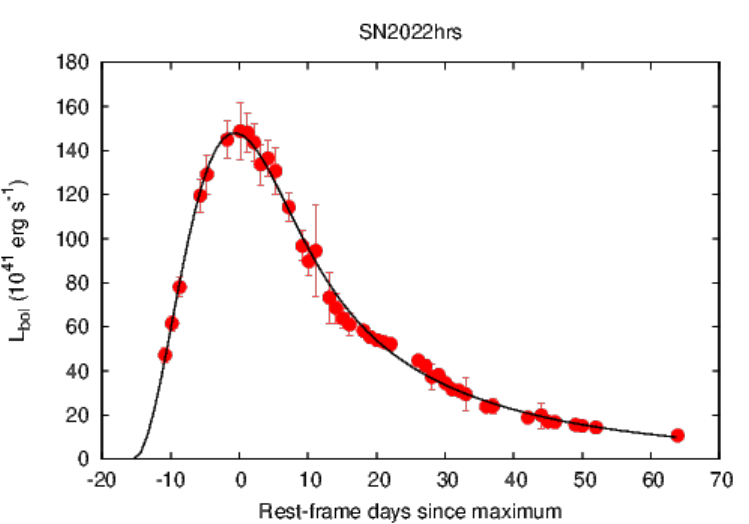}
		\includegraphics[width=0.40\linewidth]{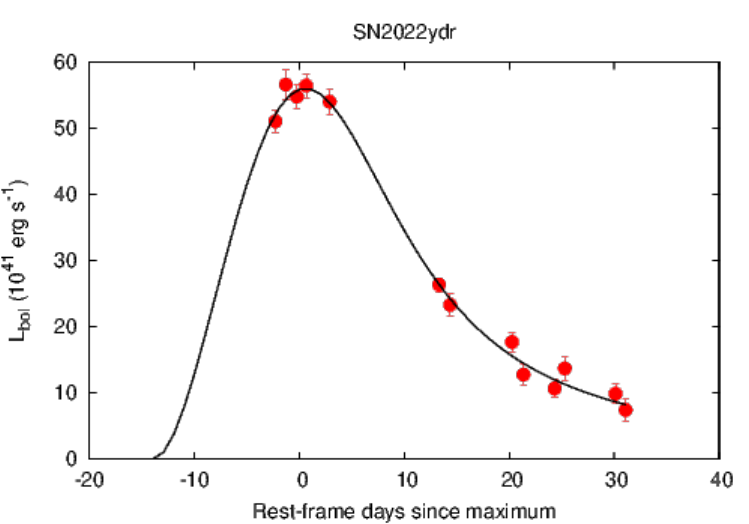}
		\includegraphics[width=0.40\linewidth]{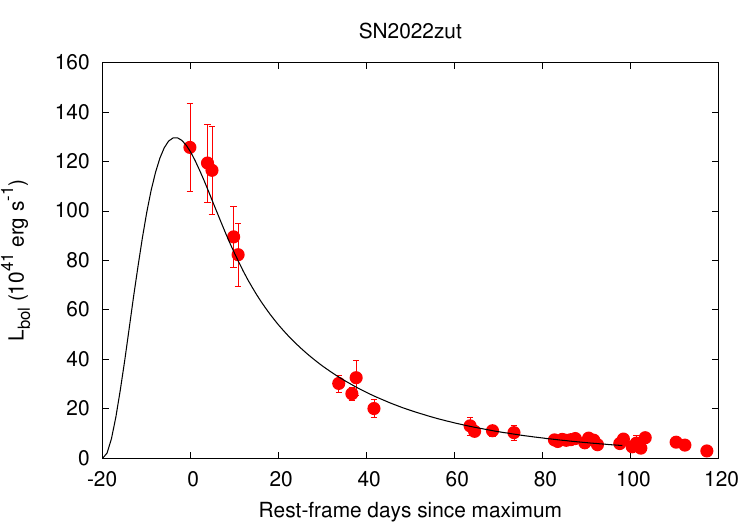}
		\includegraphics[width=0.40\linewidth]{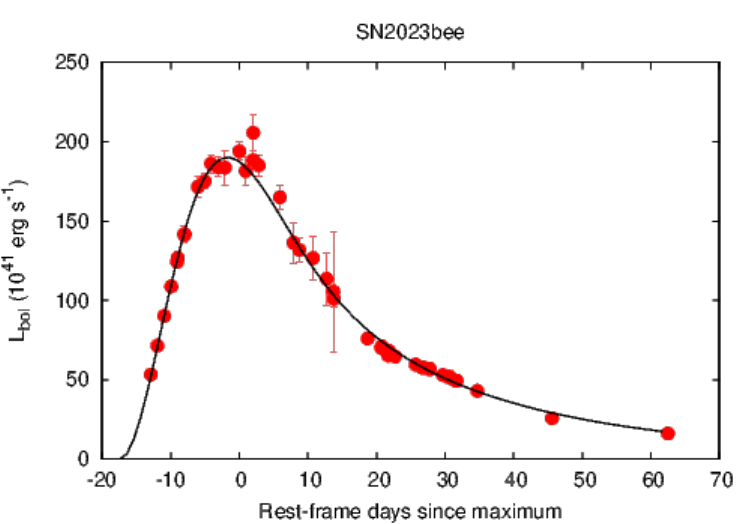}
		\caption{The same as Figure~\ref{fig:minimplots} but for additional SNe.}
	\end{figure*}

\end{document}